\documentclass[journal]{IEEEtran}
\usepackage[pdftex]{graphicx}
\usepackage{amsmath}
\usepackage{booktabs,multirow,array}
\usepackage{array}
\usepackage{makecell}
\usepackage{algorithm,algpseudocode}
\usepackage{url}
\usepackage[dvipsnames]{xcolor}
\begin{document}

% paper title
\title{Fusion of Urban TanDEM-X raw DEMs using variational models}

% author names and IEEE memberships
\author{Hossein~Bagheri,
        Michael~Schmitt,~\IEEEmembership{Senior Member,~IEEE}
       and~Xiao Xiang~Zhu,~\IEEEmembership{Senior Member,~IEEE}
\thanks{H. Bagheri and M. Schmitt are with Signal Processing in Earth Observation (SiPEO), Technical University of Munich (TUM), Germany (e-mails: hossein.bagheri@tum.de, m.schmitt@tum.de). X. Zhu is with the Remote Sensing Technology Institute (IMF), German Aerospace Center (DLR), Germany and with Signal Processing in Earth Observation (SiPEO), Technical University of Munich (TUM), Germany (e-mail: xiaoxiang.zhu@dlr.de).}
% note need leading \protect in front of \\ to get a newline within \thanks as

% \\ is fragile and will error, could use \hfil\break instead.

\thanks{The work of X. Zhu is jointly supported by the European Research Council (ERC) under the European Union's Horizon 2020 research and innovation programme (grant agreement No. [ERC-2016-StG-714087], Acronym: \textit{So2Sat}), Helmholtz Association under the framework of the Young Investigators Group ``SiPEO'' (VH-NG-1018, www.sipeo.bgu.tum.de), and the Bavarian Academy of Sciences and Humanities in the framework of Junges Kolleg.}
\thanks{Manuscript received MM DD, 200X; revised MM DD, 201X.}}

\markboth{Journal of \LaTeX\ Class Files,~Vol.~XX, No.~XX, MM~201X}%
{Shell \MakeLowercase{\textit{et al.}}: Bare Demo of IEEEtran.cls for IEEE Journals}

\maketitle

\begin{abstract}
\textcolor{Red}{This is the pre-acceptance version, to read the final version, please go to IEEE Journal of Selected Topics in Applied Earth Observations and Remote Sensing on IEEE Xplore}. Recently, a new global Digital Elevation Model (DEM) with pixel spacing of 0.4 arcseconds and relative height accuracy finer than 2m for flat areas (slopes $ < $ 20\%) and better than 4m for rugged terrain (slopes $ > $ 20\%) was created trough the TanDEM-X mission. One important step of the chain of global DEM generation is to mosaic and fuse multiple raw DEM tiles to reach the target height accuracy. Currently, Weighted Averaging (WA) is applied as a fast and simple method for TanDEM-X raw DEM fusion in which the weights are computed from height error maps delivered from the Interferometric TanDEM-X Processor (ITP). However, evaluations show that WA is not the perfect DEM fusion method for urban areas especially in confrontation with edges such as building outlines. The main focus of this paper is to investigate more advanced variational approaches such as TV-$ L_{1} $ and Huber models. Furthermore, we also assess  the performance of variational models for fusing raw DEMs produced from data takes with different baseline configurations and height of ambiguities. The results illustrate the high efficiency of variational models for TanDEM-X raw DEM fusion in comparison to WA. Using variational models could improve the DEM quality by up to 2m particularly in inner city subsets.               
\end{abstract}
\section{Introduction}
\label{sec:Introduction}
Global Digital Elevation Models (DEMs) with large coverage of the landmasses are an important source of geoinformation for different applications such as environmental studies, geographic information systems, remote sensing etc. SAR interferometry is one of the main techniques being employed for global DEM productions because of its capability to cover large areas independent of daylight or weather. For example, a global DEM with coverage of most of the planet (between 56$^{\circ}$S and 60$^{\circ}$N) was generated by Shuttle RADAR Topography Mission . The SRTM DEM is provided in the form of tiles with pixel spacings of 1 arcseconds ($\sim$ 30m) and 3 arcseconds ($\sim$ 90m)  respectively \cite{Rodriguez2006}. 
     
\begin{figure*}[ht!]
	\begin{center}
		\includegraphics[width=0.7\textwidth]{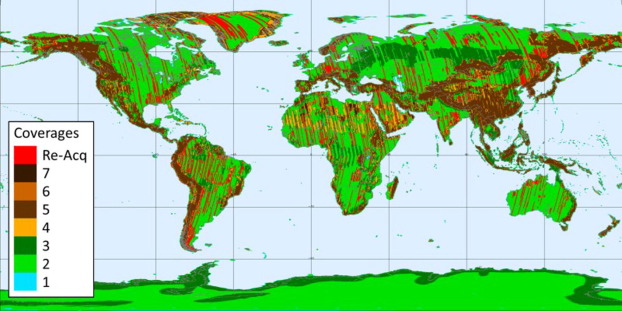}
		\caption{TanDEM-X coverage in different areas \cite{Rizzoli2017}}
		\label{tdxcoverage}
	\end{center}
\end{figure*}

Recently, a new global DEM with even higher resolution (namely a pixel spacing of 0.4 arcseconds) covering almost the whole planet was realized by the TanDEM-X DEM mission. Again, bistatic SAR acquisitions are used as input to a SAR interferometric processing chain to produce the DEM. The primary target of the mission was to provide global DEM with relative height accuracy better than 2m for flat areas (slopes lower than 20\%) and finer than 4m for remaining steeper slopes \cite{Krieger2007a}. For this, bistatic SAR data serves as excellent data source, reducing atmospheric effects and avoiding temporal decorrelation in the InSAR process. Form raw SAR data takes to the final global DEM, a workflow including different phases such as interferogram generation, phase unwrapping, data calibration, DEM block adjustment, and mosaicking is implemented at DLR \cite{Rizzoli2017}. A main step of the DEM generation procedure is carried out in the Integrated TanDEM-X Processor (ITP) which leads to primary raw DEMs for each bistatic acquisition \cite{6351133}. During the raw DEM generation, some potential error sources are removed by instrument and baseline calibration \cite{Gonzalez2012}. After that, the vertical bias which usually lies between 1m to 5m is corrected by a least squares block adjustment \cite{Gruber2012}. The block adjustment is performed by using ICESat data and connecting points in the overlapping areas of raw DEM tiles. However, dependent on the terrain morphology, some error sources still remain after the block adjustment. The effect of these errors can be decreased through fusion of several DEM coverages within the DEM Mosaicking Processor (DMP) \cite{Wessel2008}. The TanDEM-X raw DEM coverage over different terrain types is displayed in Fig. \ref{tdxcoverage}. As can be seen, the most of the world is covered by at least two nominal acquisitions with height of ambiguities (HoA) between 30m and 55m. The main objective of TanDEM-X DEM fusion is to improve the final accuracy by employing several coverages over different areas \cite{7109106}.        

Diverse methods have been designed for the fusion DEMs with different properties which can be seen as an application of data fusion in remote sensing \cite{Schmitt2016a}. Among them, Weighted Averaging (WA) is a well-established as simple approach with low computational cost \cite{7109106,7112083,Reinartz2005,dlr60015}. However its performance strongly depends on the weights that describe the height error distribution for each pixel \cite{roth2002towards}. For SAR interferometric-derived DEMs, the weights can be achieved from Height Error Maps (HEM) \cite{knopfle1998mosaicking} which are a by-product of the InSAR process and derived from the coherence values and the given geometrical configuration \cite{Just1994}. However, it should be noted that HEMs can not represent all error sources as they do not reflect deterministic effects such as layover and shadow effects. Another way is to compute weight maps by a comparison  with ground truth data that is not necessarily available for every arbitrary study area \cite{Bagheri2018}.

A current approach for implementing the DEM fusion in DMP is WA. In addition to WA, some logic for clustering consistent heights and upgrading weights regarding the influences of other significant factors such as HoA, Phase Unwrapping (PU) methodology and pixel locations relative to the border of the DEM scene is considered to finally reach the target relative accuracy and minimize PU errors remaining from primary steps \cite{7109106}. While the weighted averaging approach can realize the predefined goals in the DMP for global DEM generation, it does not perform optimally in difficult terrains with complex morphology such as urban areas which contains many high-frequency contents such as edges. After WA-based TanDEM-X DEM fusion, visualization shows that outlines of buildings are not perfectly sharp and still some amount of existing noise spoils the footprints of buildings (for example see Fig. \ref{fig:3dfusion} c). As Fig. \ref{tdxcoverage} illustrates, most areas are only covered by two nominal acquisitions (shown in green) Since this holds for many important urban areas as well, this also motivates the development of more sophisticated approaches.

An advanced approach for DEM fusion was proposed by Papasaika et al. \cite{Papasaika:2011:FDE:2050390.2050409}. They exploited sparse representations for multi sensor DEM fusion.  Zach et al. implemented an $L_{1}$ norm total variational model for range image fusion \cite{4408983}. In another study Pock et al. proposed to use Total Generalized Variation (TGV) for fusing DEMs derived from airborne optical imagery \cite{Pock2011}. Kuschk et al. evaluated weighted TGV to fuse DEMs derived from space borne optical imagery with different resolutions \cite{Kuschk2017}. In another study, weighted TV-$ L_{1} $ in which weights were predicted by neural networks were applied for the Cartosat-1 and TandEM-X DEM fusion over urban areas \cite{bagheri2017fusion}.  Overall, In spite of the high computational cost of advanced methods for DEM fusion, they perform more efficient than simple WA.       
         
In this paper, we will investigate the application of more sophisticated DEM fusion approaches that are able to efficiently preserve edges and outlines of buildings while still reducing noise effects. For this purpose, two variational models, namely  $ L_1 $ norm total variation (TV-$ L_1 $) and Huber model are implemented. Apart from these regularization approaches, we will also investigate the potential of employing raw DEMs with different properties such as different baseline configuration for the TanDEM-X DEM fusion. Therefore, this paper is structured into several sections. In section \ref{sec:Methodology}, the methodology of DEM fusion based on regularization methods is explained. Then, the description of the study subsets and experimental results from DEM fusion are provided in Section \ref{sec:exper}. Finally, the performance of the implemented DEM fusion methods for TanDEM-X data over urban areas will be discussed in Section \ref{sec:dis}.

\section{Methods for TanDEM-X Raw DEM Fusion}
\label{sec:Methodology}

In this paper, two approaches are implemented for TanDEM-X raw DEM fusion. The characteristics of each model will be explained in the following.
Before the fusion, raw DEMs at first are aligned to each other by DEM coregistration approaches such as least square matching \cite{GRUEN2005151}, iterative closest point \cite{Ravanbakhsh2013} or manual registration. The coregistration of DEMs decreases their translational and rotational differences. For stability reasons, in addition, the height data should be normalized to the interval [0, 1]\cite{Kuschk2017}: 

			\begin{equation}\label{normal}
			h_{k}^{n}{(x,y)}=\dfrac{h_{k}{(x,y)}-h_{min}}{h_{max}-h_{min}}
			\end{equation}
			
where $ h_{k}{(x,y)>0} $ is the elevation of the study DEM with index $k$ at location $(x,y)$, $h_{max}>0$ and $h_{min}>0$ ($ h_{min}< h_{max} $) are the lowest and highest elevations among all input DEMs. The output gives the normalized height in the considered location.       		

\subsection{Background: Weighted Averaging}
\label{ssec:WA}

 The most popular, very fast and low computational cost method for DEM fusion is weighted averaging which is implemented by

	        \begin{equation}\label{WA}
	    	\mathbf{f}={\sum_{i=1}^{k}{\mathbf{w}_{i}} \odot {\mathbf{h}_{i}}}
	    	\end{equation}
	    	where $\mathbf{h}_{i} $ are 2D arrays representing the input DEMs, $\mathbf{w}_{i}$ are the corresponding weight maps and $\odot$ is a pixel-wise product. It is worth to note that other simple methods such as pixel-wise median or mode based fusion can also be employed for DEM fusion especially when multiple DEMs are available \cite{Rumpler2013}. 
	    	 
As explained in Section \ref{sec:Introduction}, the main critical issue for using weighted averaging for DEM fusion is to apply appropriate weights that are fairly representative of expected height errors in the source DEMs.  For TanDEM-X DEM fusion, generally, these weights are delivered as Height Error Maps (HEMs) from the ITP. For each height of the TanDEM-X DEM, the corresponding HEM value can be estimated by:
	    	
	    	\begin{equation}\label{HE}
	    	{\sigma}_{j}=H_{amb}\,\,\frac{{\sigma}_{\phi,j}}{2\pi}
	    	\end{equation}
	    	where $H_{amb}$ is the height of ambiguity and ${\sigma}_{\phi,j}$ is the interferometric phase error that is estimated from the interferometric coherence and the InSAR geometry \cite{Krieger2007a}. Then, from these values, the respective weights can be calculated for each pixel location by 
	    	
 	    	\begin{equation}\label{weight}
 	    	{w}_{j}=\frac{\frac{1}{{\sigma}_{j}^{2}}}{\sum_{j=1}^{N}\frac{1}{{\sigma}_{j}^{2}}}
 	    	\end{equation}

\subsection{Regularization-Based Models}
\label{sec:DEMfusion}

Variational models were firstly used for signal and image denoising \cite{Rudin1992a,Nikolova2004}. Generally, in variational denoising approaches, an energy functional is constituted by fidelity and regularization terms. The fidelity is considered to enforce the output image being similar to the input images while the regularization term (also called penalty term) is embedded to reduce the effect of noise in the final result. The desired output is achieved by minimizing the constructed energy functional. Diverse energy functionals can be formed according to different functions for defining data and penalty terms \cite{Pock2011}. 

A popular type of variational models is the total variation-based model (TV) in which the gradient of desired output image is selected to form the regularization term based on different norms. The main advantage of the TV-based variational model its convexity that guarantees to find a solution by minimizing the energy functional.

In the problem of TanDEM-X DEM fusion, several input raw DEMs are fused using variational models. The data term makes the fused DEM similar to the input tiles while the TV-based regularization term is defined to provide a sharp output at the end by preserving the edges and reducing the noise. This property is beneficial for fusing TanDEM-X raw DEMs over urban areas where footprints of buildings as edges often appear very noisy because of the inherent SAR imaging properties.         
 
The basic gradient-based variational model for image denoising and data fusion is a quadratic model in which $ L_2 $ norm is used for both regularization and data terms \cite{tikhonov1943stability}. However the quadratic regularization term causes over-smoothing for edges. Therefore, using the $ L_1 $ norm instead was proposed by Rudin, Osher, and Fatemi which is called ROF model correspondingly \cite{Rudin1992a}. Since the ROF model still uses the $ L_2 $ norm for the data term, it does not provide robustness against outliers when applied to DEM fusion. As a solution, the $ L_1 $ norm can be substituted for the $ L_2 $ norm \cite{doi:10.1137/040604297}. The TV-$ L_1 $ model consists of the data fidelity and the penalty term: 
 
		\begin{equation}\label{tv}
		\min_\mathbf{f}\bigg\{\sum_{i=1}^{k}\Arrowvert \mathbf{f}-\mathbf{h}_{i}\Arrowvert_{1}+\gamma\Arrowvert \nabla \mathbf{f}\Arrowvert_{1} \bigg\}
		\end{equation}	
	where $\mathbf{h}_{i}$ are noisy input  DEMs and $\mathbf{f}$ is the desired DEM should be achieved by minimizing the functional energy above. The penalty term is formed based on the gradients of the newly estimated DEM to preserve the edges at the end. The regularization parameter $\gamma$ trades off between penalty and fidelity terms. Increasing $\gamma$ will influence the smoothness and will produce a smoother fused DEM in the end.     

While the main advantage of TV-$ L_1 $ is its robustness against strong outliers as well as edge preservation \cite{Pock2011}, it suffers from the staircasing effect, a phenomenon that creates artificial discontinuities in the final output and particularly affects high resolution DEM fusion \cite{Chan2006}. Moreover, the $ L_1 $ norm is not necessarily the best choice for all data fusion and denoising cases. As an alternative, the Huber regularization model is proposed to rectify the drawbacks of the TV-$ L_1 $ model \cite{Pock2011}. It applies the Huber norm instead of the $ L_1 $ norm in both fidelity and penalty terms \cite{Huber2011}: 

		\begin{equation}\label{hubernorm}	
		\Arrowvert {x}\Arrowvert_{\eta}=\begin{cases} \frac{\lvert {x}\lvert^{2}}{2\eta}& \text{if $\lvert {x}\lvert\leq\eta$}.\\ \lvert {x}\lvert-\frac{\eta}{2} & \text{if $\lvert {x}\lvert > \eta$}. \end{cases}
		\end{equation} 
		Here, $\eta$ is a parameter that determines a threshold between the $ L_1 $ and $ L_2 $ norm in the model. Based on this, the Huber model can be defined as \cite{doi:10.5721/EuJRS20164926}

		\begin{equation}\label{huber}
		\min_\mathbf{f}\bigg\{\sum_{i=1}^{k}\sum_\Omega \Arrowvert \mathbf{f}-\mathbf{h}_{i}\Arrowvert_{\alpha}+\gamma\sum_\Omega \Arrowvert \nabla \mathbf{f}\Arrowvert_{\beta} \bigg\}
		\end{equation} 
		where both data and penalty terms are constituted based on the thresholds $\alpha$ and $\beta$ that are substituted as $ \eta $ in the Huber norm relation (\ref{hubernorm}) to form these terms and $ \Omega $ denotes the raster DEM space. It should be noted that the Huber norm is a generalized form of the $ L_1 $ norm. However, in this study the Huber norm is also used to strictly penalize the outliers.
%%%
%	\begin{figure}[htb]
%	\begin{minipage}[b]{1\linewidth}
%		\centering
%		\centerline{\epsfig{figure=norm,width=0.9\columnwidth, height=0.7\columnwidth}}
%		\vspace{0.05cm}		
%	\end{minipage}
%	\caption{Different norms used to form diverse TV-based variational models for fusion of TanDEM-X raw DEMs. Blue: Huber norms, Red: quadratic norm, and Black dashed: $ L_1 $ norm.}
%	\label{fig:norm}
	%
% \end{figure} 
%%%
		
%Figure \ref{fig:norm} displays different norms that are used in the variational models such as the quadratic norm (the red curve), Huber norm with different values of $ \eta $  (blue curves) and $ L_1 $ norm (the black dashed plot). The plot illustrates that%
 Using quadratic norm in the regularization term penalizes high frequency changes more than $ L_1 $ norm and thus, it reduces the noise at the cost of oversmoothing edges. The Huber norm, dependent on $ \eta $  values treats as a norm between $ L_1 $ and $ L_2 $ norms. however for $ \eta = 1$, its behavior is nearly similar to $ L_1 $. In other words, the Huber norm with higher $ \eta $ provides DEMs with smoother building footprints but not as much as the quadratic norm. The influence of the parameters of variational models on the quality of fused DEM will be discussed in Section \ref{sec:dis-model} with more details. In the remainder of this paper, we use $ \alpha=4 $ to smooth relative height errors larger than 4m (considering the relative accuracy of the TanDEM-X DEM), and $ \beta=1 $ based on data driven experiments on different datasets. Consequently, $ \gamma $ can be calculated by L-curve method \cite{doi:10.1137/0914086}.   
							
\subsection{Implementation}
It is mathematically proven that the TV-based energy functional based on $L_1 $ or Huber norm is convex. The main characteristic of a convex problem is that the desired output (i.e. the global minimum) will be certainly found through an optimization process. One popular strategy for finding the minimum of a convex optimization is to reformulate the functional energy as primal-dual problem \cite{Chambolle2011}. For variational models, the energy functional can be expressed in a general form such as

	\begin{equation}\label{pCVX}
	\min_\mathbf{u}\big\{\mathcal{G}(\mathbf{u})+\mathcal{F}(K\mathbf{u}) \big\}
	\end{equation} 
where $ \mathcal{G}(\mathbf{u}) $ is the data term and $ \mathcal{F}(K\mathbf{u})  $ is the regularization term. $ K $ refers to an operator that is used for defining the regularization term (for TV-based variational models, this is the gradient $ \nabla $ of the desired output). In the TanDEM-X raw DEM fusion, $ \mathbf{u} $ (as primal variable) is the final fused DEM ($ \mathbf{f} $) and the energy functional (terms $ \mathcal{G}(\mathbf{u}) $ , $ \mathcal{F}(K\mathbf{u})  $) is defined by relations (\ref{tv}), and (\ref{huber}).  		
Then, the dual-problem formulation of the energy functional can be written as:
 	
	\begin{align}\label{dual}
	\min_\mathbf{u}\max_\mathbf{v}\big\{\mathcal{G}(\mathbf{u})+\langle \mathbf{v},K\mathbf{u}\rangle-\mathcal{F^*}(\mathbf{v})\big\} \\
	\textrm{where} \quad \mathcal{F^*}(\mathbf{v}^*)=\sup_{\mathbf{v} \in \mathcal{\mathbf{V}}}\langle \mathbf{v}^*,\mathbf{v} \rangle-\mathcal{F}(\mathbf{v})
	\end{align} 
and $ \mathbf{v} $ is the dual variable and $\mathcal{F^*}$ is defined as the convex conjugate of $\mathcal{F}$ . The dual-problem algorithm for minimizing equation (\ref{dual}) is presented in algorithm \ref{dpalgorithm}. It should be noted that the median-based fusion of the input DEMs can be used to initialize $\mathbf{u}^0$ to speed up the optimization. More details of the algorithm can be found in \cite{Chambolle2011}.    
		
	\begin{algorithm}
		\caption{Dual primal algorithm}
		\label{dpalgorithm}
		\algnewcommand\algorithmicinput{\textbf{Input:}}
		\algnewcommand\Input{\item[\algorithmicinput]}
		\algnewcommand\algorithmicoutput{\textbf{Output:}}
		\algnewcommand\Output{\item[\algorithmicoutput]}
		\begin{algorithmic}[1]
			\Input{Primary DEMs to configure primal problem}
			\State Initialization: $\tau\sigma\Arrowvert K\Arrowvert \le1$ , $(\hat{\mathbf{u}}^0,\mathbf{v}^0)\in \mathcal{\mathbf{U}}\times \mathcal{\mathbf{V}}$,$\hat{\mathbf{u}}^0 = \mathbf{u}^0$, $\theta\in[0,1]$, 
			\For{$i=0$ to stopping criteria}
			\begin{align*}
			\mathbf{v}^{i+1}&=(I+ \sigma \partial \mathcal{F^*})^{-1}(\mathbf{v}^i+\sigma K \hat{\mathbf{u}}^i)  \\
			\mathbf{u}^{i+1}&=(I+\tau \partial \mathcal{G})^{-1}({\mathbf{u}}^i-\tau K^T\mathbf{v}^{i+1})   \\
			\hat{\mathbf{u}}^{i+1} &= \mathbf{u}^{i+1}+\theta (\mathbf{u}^{i+1}-\mathbf{u}^{i})
			\end{align*}
			\EndFor
			\Output{$\mathbf{u}$}
		\end{algorithmic}
	\end{algorithm}
	
\section{Experiments}\label{sec:exper}
In this paper, we investigate TanDEM-X raw DEM fusion over urban areas by using diverse TV-based variational models such as TV-$ L_{1} $ and Huber models. In addition, the effect of fusing raw DEMs with different baseline configurations will be investigated. The  baseline configurations for TanDEM-X data differ by changing orbit direction (Ascending or Descending) and also changing HoA values. Furthermore, the results of DEM fusion implementation will be evaluated for different land types with an emphasize on urban areas. These land types are: 1) industrial areas that are characterized as areas with large buildings, often not very high, 2) inner city areas that include very densely packed buildings, relatively high and 3) residential areas that are typically specified with low rise single family homes. In addition, we also considered some non-urban study areas such as agricultural and forested areas to evaluate the performance of variational models in those areas too. 

In TanDEM-X raw DEMs produced with a pixel spacing of 0.2 arcseconds (around 6m), most building footprints in industrial and inner city areas can be visualized but the height accuracy and quality of building shapes suffer from noise and systematic errors such as layover and shadow. The visualization and quality of building become worse in residential areas because of the small sizes and heights of buildings in these areas. However, we will evaluate the performance of variational models for enhancing the quality of buildings appeared in the final fused DEM over different aforementioned land types. After resampling and coregistration, the raw DEMs are fused by the different approaches explained in Section \ref{sec:Methodology}.

\subsection{Fusion of TanDEM-X Raw DEMs with Similar Baseline Configuration}\label{sec:similarbase}

Most of the global coverage achieved with the TanDEM-X raw DEMs is generated by two nominal bistatistic acquisition (see Fig. \ref{tdxcoverage}), but there are more tiles in overlapping areas at the border of the tiles. The first investigation includes data takes that have similar baseline configurations as well as HoAs. The study subsets are selected from two nominal TanDEM-X raw DEMs over Munich city in Germany. The characteristics of these raw DEMs are presented in Tab \ref{PropertyTDX_munich}. 

\begin{figure}[htb]
	\begin{minipage}[b]{0.49\linewidth}
		\centering
		\centerline{\includegraphics[width=1\columnwidth]{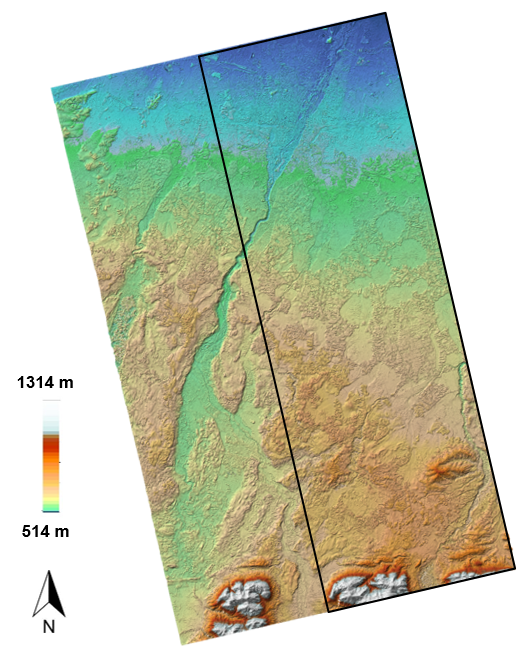}}
		\vspace{0.05cm}
		\centerline{(a)}\medskip
	\end{minipage}
	\begin{minipage}[b]{0.49\linewidth}
		\centering
		\centerline{\includegraphics[width=1\columnwidth]{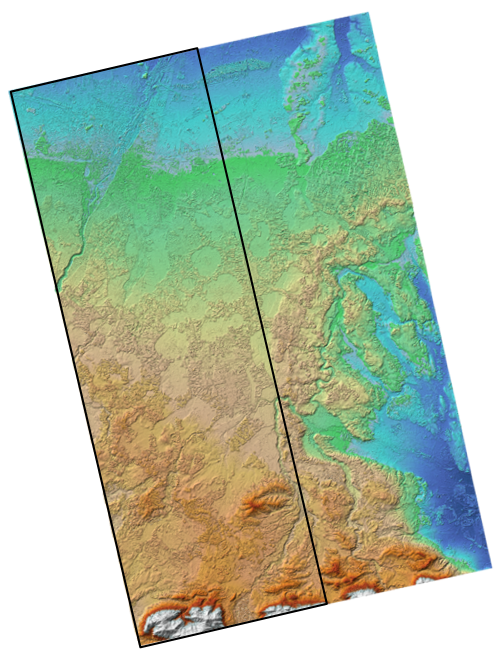}}
		\vspace{0.05cm}
		\centerline{(b)}\medskip
	\end{minipage}
	\caption{Two nominal acquisitions of TanDEM-X raw DEMs over Munich area. (a) 1145180 tile and (b) 1023491  tile}
	\label{fig:TDXtiles}
\end{figure}

\begin{figure}[htb]

		\centering
		\centerline{\includegraphics[width=1\columnwidth]{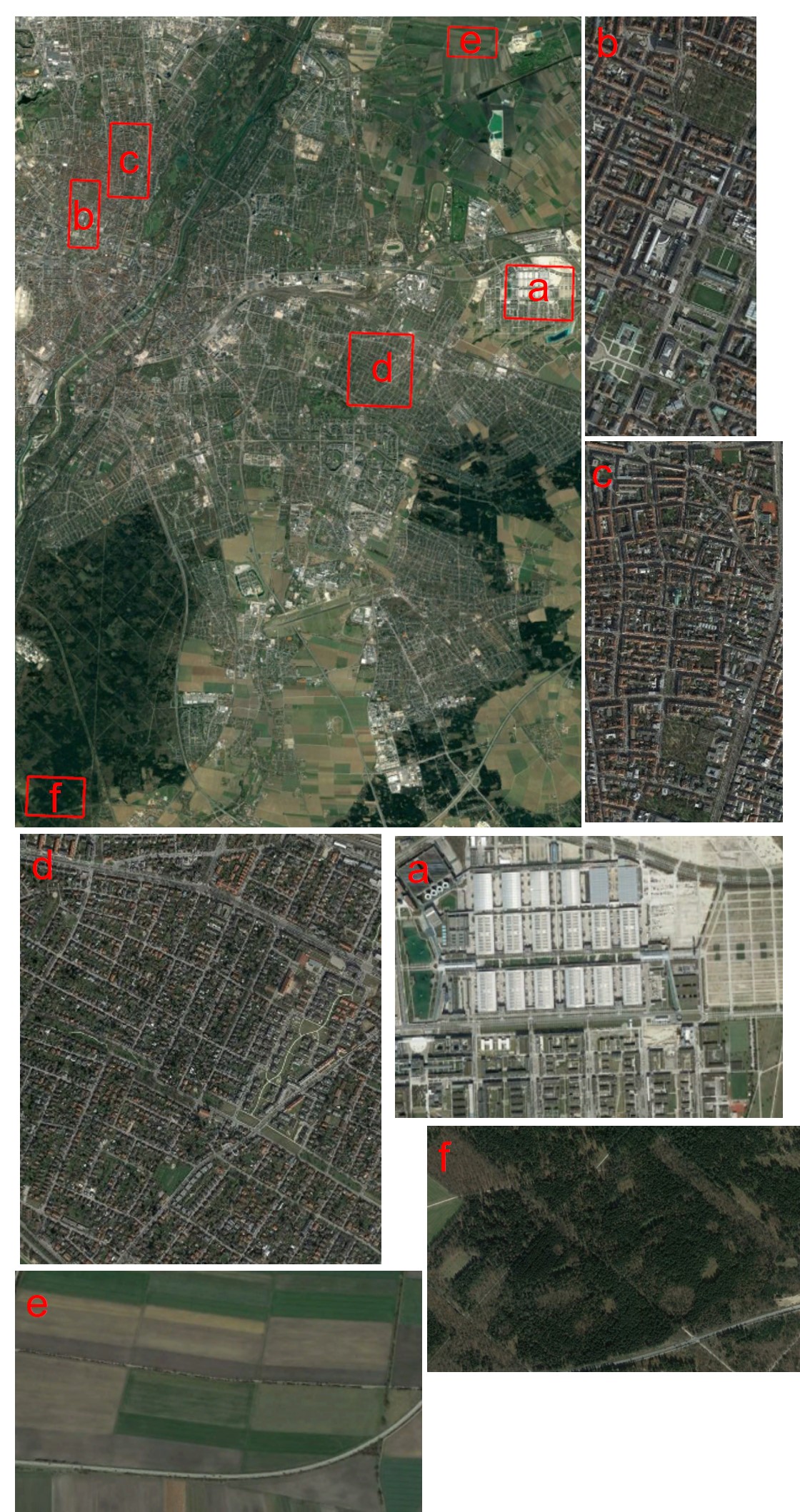}}
		\vspace{0.05cm}
	\caption{ A display of the study subsets selected from different urban land types for TanDEM-X raw DEM fusion over Munich city; a) industrial area (1.5km $ \times $ 1.2km), b) inner city 1 (1.5km $ \times $ 0.6km), c) inner city 2 (1.6km $ \times $ 0.9km),  d) residential area (1.6km $ \times $ 1.3km), e) agricultural (1.05km $ \times $ 0.6km), and f) forested (1.25km $ \times $ 0.85km)}
	\label{fig:subset1}
\end{figure}

Figure \ref{fig:TDXtiles} displays the raw DEM tiles used for this experiment. From those, 4 subsets as representatives of different land types are extracted for the DEM fusion task. A display of these subsets is provided in Fig. \ref{fig:subset1}.

 \begin{table}[h]
	\centering
	\caption{Properties of the nominal TanDEM-X raw DEM tiles for Munich area}
	%\resizebox{\columnwidth}{!}{%
	\begin{tabular}{l c c }
		\hline
		\multicolumn{3}{c}{TanDEM-X raws DEMs: Munich area}\\
		Acquisition Id                 & 1023491                & 1145180                     \\\hline
		Acquisition mode      & Stripmap  & Stripmap \\
		Center incidence angle      & 38.25$^{\circ}$  & 37.03$^{\circ}$  \\
		Equator crossing direction  & Ascending        & Ascending          \\
		Look direction              & Right            & Right                  \\
		Polarization                & HH               & HH                        \\
		Height of ambiguity         & 45.81m           & 53.21 m                \\
		Pixel spacing               & 0.2 arcsec       & 0.2 arcsec        \\
		HEM mean                    & 1.33 m           & 1.58                   \\
		\hline		
	\end{tabular} 
	%} 
	
	\label{PropertyTDX_munich}
\end{table}


The quality of the input raw TanDEM-X DEMs as well as the fused version are determined by using the reference LiDAR DEM which is produced from a high resolution airborne LiDAR point cloud acquired over Munich and provided by Bavarian Surveying Administration. The density of the LiDAR point cloud changes for each subset, but at least there is one point per square, and the vertical accuracy of the point cloud is better than $ \pm $ 20cm. The final reference DEM is achieved by interpolation in a grid with pixel spacing as same as input TanDEM-X DEMs.       

The results of raw DEM fusion using  TV-$L_{1} $ and Huber models for study areas are presented in Tab \ref{tab:result1}. The regularization parameter is calculated using L-curve method. For comparison with the common fusion method, the results of fusion by WA are also provided. The DEM quality after and before fusion was evaluated by statistical metrics, Mean, Root Mean Square Error (RMSE), Mean Absolute Error (MAE), Normal Median Absolute Deviation (NMAD), and Standard Deviation (STD).   

	\begin{table*}[h]
	\centering
	\caption{Height accuracy (in meter) of the TanDEM-X data before and after DEM fusion in the different study areas over Munich.}
	%\resizebox{\columnwidth}{!}{%
	\begin{tabular}{lll|ccccc}\hline
		
		Study area&\multicolumn{2}{c|}{DEM}& Mean & RMSE & MAE & NMAD & STD \\
		\hline
		\multirow{5}{*}{Industrial}  		
		&\multirow{2}{*}{Raw DEM}&id: 1023491 & 0.71 &	4.40 &	3.08 &	2.37 &	4.34  \\
		&&id: 1145180  & 0.71 &	4.64 &	3.27 &	3.01 &	4.58  \\                         
		&\multirow{3}{*}{Fused DEM} & WA   & 0.77 &	4.16 &	2.93 &	2.24 &	4.09  \\
		&& TV-$ L_1 $	     & \textbf{0.69} &	\textbf{3.67} &	\textbf{2.69} &	\textbf{2.03} &	\textbf{3.60}  \\
		&& Huber	      & 0.71 &	3.74 &	2.84 &	2.40 &	3.67  \\
		\hline
		\hline
		\multirow{5}{*}{Inner 1}
		&\multirow{2}{*}{Raw DEM}   &id:1023491    &0.78&	7.79&	5.95&	6.49&	7.75 \\
		&                           &id:1145180    &0.78&	8.08&	6.30&	7.15&	8.04 \\
		&\multirow{3}{*}{Fused DEM} & WA        &0.84&	7.51&	5.83&	6.49&	7.46 \\
		&                           & TV-$ L_1 $& \textbf{0.77} &	\textbf{6.11} &	\textbf{5.00} &	5.72 &	\textbf{6.06} \\
		&                           & Huber	    & 0.78 &	6.14 &	5.09 &	\textbf{5.67} &	6.09 \\
		\hline
		\hline
		\multirow{5}{*}{Inner 2}
		&\multirow{2}{*}{Raw DEM}   &id:1023491    &0.18&	7.00&	5.44&	6.36&	7.00 \\
		&                           &id:1145180    &0.18&	7.16&	5.57&	6.51&	7.16 \\
		&\multirow{3}{*}{Fused DEM} & WA        &0.20&	6.82&	5.33&	6.23&	6.82 \\
		&                           & TV-$ L_1 $& \textbf{0.12} &	5.83 &	\textbf{4.78} &	\textbf{6.16} &	5.83  \\
		&                           & Huber	    & 0.18 &	\textbf{5.82} &	4.82 &	6.23 &	\textbf{5.82}  \\
		\hline
		\hline
		\multirow{5}{*}{Residential}
		&\multirow{2}{*}{Raw DEM}   &id:1023491    &0.95&	2.68&	2.10& 	2.05&	2.50 \\
		&                           &id:1145180    &0.95&	2.92&	2.25&	2.31&	2.76 \\
		&\multirow{3}{*}{Fused DEM} & WA        &0.96&	2.61&	2.05&	1.99&	2.43 \\
		&                           & TV-$ L_1 $& \textbf{0.89} &	\textbf{2.41} &	\textbf{1.96} &	\textbf{1.98} &	\textbf{2.24}  \\
		&                           & Huber	    & 0.95 &	2.44 &	1.98 &	\textbf{1.98} &	\textbf{2.24}  \\
		\hline
		 \hline		
		\multirow{5}{*}{Agricultural}		
		&\multirow{2}{*}{Raw DEM}   &id:1023491    &0.13	&0.86	&0.57	&0.59	&0.84 \\	
		&                           &id:1145180    &0.13	&1.64	&1.13	&1.20	&1.64 \\	
		&\multirow{3}{*}{Fused DEM} & WA        &0.14	&0.78	&0.51	&0.54	&0.76 \\	
		&                           & TV-$ L_1 $&\textbf{0.06}	&\textbf{0.55}	&\textbf{0.29}	&\textbf{0.20}	&\textbf{0.54} \\
		&                           & Huber	    &0.13	&0.72	&0.48	&0.47	&0.71 \\	
		\hline
		\hline
		\multirow{5}{*}{Forested}
		&\multirow{2}{*}{Raw DEM}   &id:1023491    &\textbf{2.25}	&4.84	&3.54	&3.46	&4.28  \\
		&                           &id:1145180    &\textbf{2.25}	&4.58	&3.36	&3.24	&3.99  \\
		&\multirow{3}{*}{Fused DEM} & WA        &2.28	&4.51	&3.30	&3.17	&3.89  \\
		&                           & TV-$ L_1 $&\textbf{2.25}	&\textbf{4.34}	&\textbf{3.18}	&\textbf{3.09}	&\textbf{3.71}  \\
		&                           & Huber	    &\textbf{2.25}	&4.36	&3.21	&3.12	&3.73	\\
		\hline
	\end{tabular}
	%	}
	
	\label{tab:result1}
\end{table*}

\begin{figure*}[htb]
	\centering
	\begin{minipage}[b]{1\linewidth}
		\centering
		\centerline{\includegraphics[width=1\columnwidth]{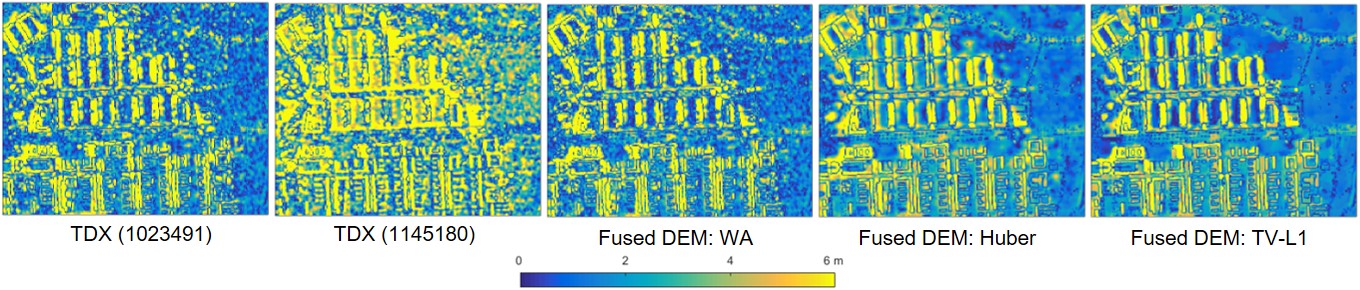}}
		\vspace{0.05cm}
		\centerline{(a)}\medskip
	\end{minipage}

	\centering
	\begin{minipage}[b]{0.55\linewidth}
		\centering
		\centerline{\includegraphics[width=1\columnwidth]{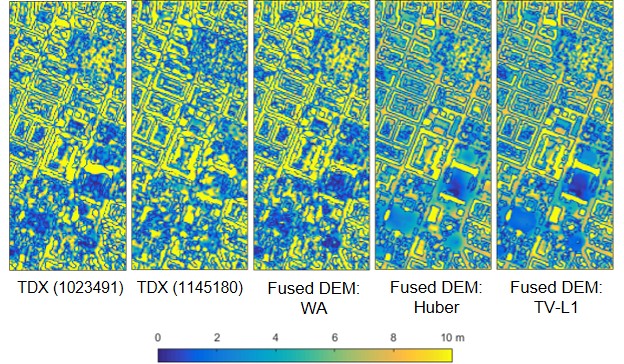}}
		\vspace{0.05cm}
		\centerline{(b)}\medskip
	\end{minipage}
	\caption{Absolute residual maps of the initial input raw DEMs and the fused DEMs obtained by different approaches for the industrial (a) and inner city (b) study areas over Munich }
	\label{fig:residmap}
\end{figure*}
In addition to statistical analysis, to evaluate the performance of variational models, the residual maps of the input DEMs and the fused DEMs achieved by different methods for the industrial and inner city 1 study areas are displayed in Fig \ref{fig:residmap}.  
The results illustrate using variational models in the fusion process can finally improve the quality of the TanDEM-X DEM over the quality achievable with classic WA. It is explicitly displayed on residual maps that variational modes can finally reduce the noise effects and also makes the footprints of buildings more apparent than WA. Furthermore, fusing ascending and descending DEMs can improve the DEM quality in particular for the shadow- and layover-affected areas in which significant errors occur. 

\subsection{Fusion of TanDEM-X Raw DEMs with Different HoAs}\label{sec:diffHoA}
In the first experiment, the study areas were selected from two TanDEM-X raw DEMs that have nearly similar properties. Both DEMs were acquired in the same orbit, same look directions and also with nearly the same incidence angles and HoAs.      
  
\begin{figure}[htb]
	\centering
	\begin{minipage}[]{0.7\linewidth} 
		\centering
		\centerline{\includegraphics[width=1\columnwidth]{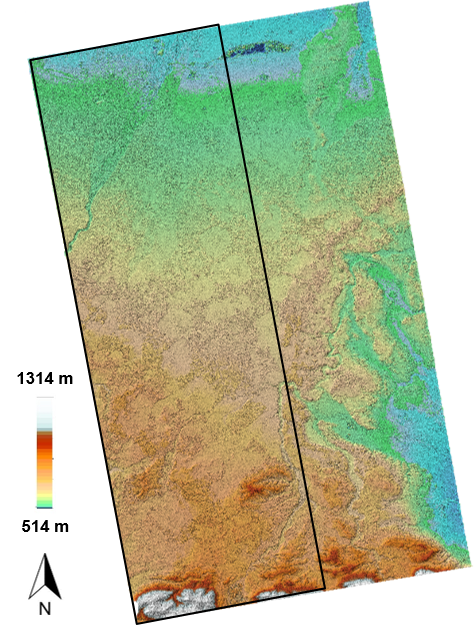}}
	\end{minipage}
	\caption{The non-official TanDEM-X raw DEM tile produced with bigger HoA over Munich area}
	\label{fig:TDXtilec}
\end{figure}

As an additional experiment, we investigate the performance of variational models for fusing TanDEM-X raw DEMs with different HoAs over urban areas. For this purpose, one experimental ITP raw DEM with different HoA over Munich city in Germany is considered. It should be noted this product has not been used in final global DEM generation but in this study is applied for implementing an experiment of fusing rawDEMs with different HoAs. The specifications of this raw DEM are shown in Tab. \ref{PropertyTDX}. Figure \ref{fig:TDXtilec} also provides a depiction of the new raw DEM which is acquired over the same location as tile 1023491 with identical overlap.      
  
The main property that discriminates this tile from those introduced in the previous section is its bigger HoA. Regarding nearly similar incidence angle and slant range, the larger value for HoA means this tile is derived from data takes that were acquired with a shorter baseline is considered helpful in areas where phase unwrapping (PU) errors are dominant \cite{Martone2012}. On the other hand, the quality and resolution of this DEM is lower than those with smaller HoAs. Comparing the raw tiles displayed in Figs. \ref{fig:TDXtiles} and \ref{fig:TDXtilec} confirms, a drop of quality of DEM, i.e. with much more noise, with id 1058842 which has larger HoA. 

\begin{table}[h]
	\centering
	\caption{Properties of the non-official TanDEM-X raw DEM tile for Munich area}
	%\resizebox{\columnwidth}{!}{%
	\begin{tabular}{l   c}
		\hline
		\multicolumn{2}{c}{TanDEM-X raws DEMs: Munich area}\\
		Acquisition Id                 & 1058842        \\\hline
		Acquisition mode      & Stripmap \\
		Center incidence angle      & 38.33 $^{\circ}$ \\
		Equator crossing direction  &  Ascending   \\
		Look direction              &   Right       \\
		Polarization                &  HH          \\
		Height of ambiguity         &  72.02       \\
		Pixel spacing               &  0.2 arcsec  \\
		HEM mean                    &  2.58        \\
		\hline		
	\end{tabular} 
	%} 
	
	\label{PropertyTDX}
\end{table}

\begin{figure}[htb]
	\centering
	\begin{minipage}[]{0.45\linewidth} 
		\centering
		\centerline{\includegraphics[width=1\columnwidth]{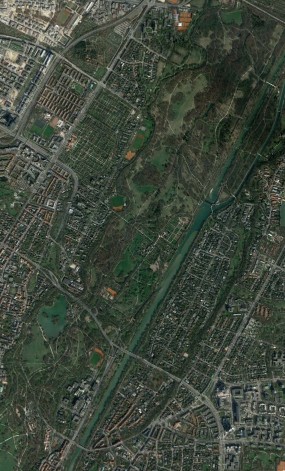}}
		\vspace{0.05cm}
	\end{minipage}
	\caption{The study subset selected for DEM fusion in a problematic area (4.5km $ \times $ 2.8km).}
	\label{fig:studysub2}
\end{figure}

In this experiment, a study subset is extracted from an area that has lots of inconsistent heights due to PU errors. For this aim, a relatively large subset from an urban area which is covered by trees and also includes a river crossing is selected. Figure \ref{fig:studysub2} displays the selected study area suffering from PU errors. The corresponding DEM data are derived from tiles  1023491 and 1058842 with HoAs about 45m and 72m respectively.

The PU errors appearing in this subset originate from the volume decorrelation phenomenon that happens in an area covered by trees (like the selected study subset)  and also a coherence change due to transition from dry land to water (river). PU errors typically are at the range of multiples of the HoA value. The inconsistent heights can be determined by \cite{7109106}: 
	\begin{equation}\label{thPU}
     dh_{th}= 0.75\times min (|HoA|)-4
     \end{equation}
Those height residuals bigger than $ dh_{th} $ are denoted as inconsistent height values emerging because of PU errors.

Table \ref{tab:result2} collects the results of fusing DEMs with different HoAs in the selected study area. Again, the accuracy was evaluated respective to a reference DSM interpolated from a point cloud with high density (more than 8 points per square meters). Moreover, Tab. \ref{tab:result3} compares the fused DEMs with different approaches and initial DEMs in terms of number of PU errors, maximum and minimum height residuals. The PU threshold for each DEM is computed based on the respective HoA value using (\ref{thPU}). It is obvious that the DEM 1058842 has lower number of PU errors because of larger HoA but for DEM fusion quality analysis, the minimum value of HoAs (here 45.81) is considered to enumerate the number of PU errors. It should be noted that mean values presented in Tabs. \ref{tab:result1} and \ref{tab:result2} do not present the real level of canopy penetration of the X-band radar signal. In our previous study \cite{isprs-archives-XLII-1-W1-433-2017}, we found some amount of vegetation penetration to remain after DEM coregistration.         

	\begin{table*}[h]
	\centering
		\caption{Height accuracy (in meter) of the TanDEM-X data with different HoAs before and after DEM fusion in the problematic study area.}
	%\resizebox{\columnwidth}{!}{%
	\begin{tabular}{ll|ccccc}\hline
		
		\multicolumn{2}{c|}{DEM}& Mean & RMSE & MAE & NMAD & STD \\
		\hline 		
		\multirow{2}{*}{Raw DEM} &id: 1023491   &\textbf{-2.35}	&10.77	&8.46	&10.10	&10.51  \\
	                           	&id: 1058842   &\textbf{-2.35}	&10.57	&8.27	&9.69	&10.30  \\
		\multirow{3}{*}{Fused DEM} & WA     &-2.37	&10.45	&8.23	&9.81	&10.17  \\
		 & TV-$ L_1 $	                     &-2.63	&9.24	&7.13	&8.03	&8.86   \\
		& Huber	                             &\textbf{-2.35}	&\textbf{8.60}	&\textbf{6.70}	&\textbf{7.65}	&\textbf{8.277}   \\
		\hline
	\end{tabular}
	%	}

	\label{tab:result2}
\end{table*}  
	\begin{table*}[h]
	\centering
	\caption{Effect of DEM fusion to reduce the number of PU errors using  tiles with different HoAs in the problematic study area.}
	%\resizebox{\columnwidth}{!}{%
	\begin{tabular}{ll|ccccc}\hline
		
		\multicolumn{2}{c|}{DEM}&HoA &PU Threshold & No. of PU Errors & Max Discrepancy & Min Discrepancy \\
		\hline 		
   	\multirow{2}{*}{Raw DEM} &id: 1023491   &45.81	&30.36	&2032	&51.80	&-73.13         \\
	                         &id: 1058842    &72.02	&50.01	&51	    &58.82	&-54.76         \\
	\multirow{3}{*}{Fused DEM} & WA     &45.81	&30.36	&1339	&50.74	&-53.39         \\
	& TV-$ L_1 $	                     &45.81	&30.36	&102	    &19.16	&-33.76     \\
	& Huber	                             &\textbf{45.81}	&\textbf{30.36}	&\textbf{0}	    &\textbf{16.97}	&\textbf{-28.71}         \\

		\hline
	\end{tabular}
	%	}
	
	\label{tab:result3}
\end{table*}  

 The results from Tab. \ref{tab:result2} and \ref{tab:result3} demonstrate the efficiency of the Huber model for fusion of two tiles of TanDEM-X raw DEMs in the problematic area. The results show that using the Huber model can significantly improve the RMSE of fused DEMs by up to nearly 2m while the DEM quality enhancement by means of WA is not remarkable. Apart from this, the Huber model is absolutely more powerful than WA to reduce the PU errors. The maximum and minimum discrepancies also confirm the better performance of the Huber model to deal with PU errors in comparison to the other method. TV-$ L_1 $ also can decrease the noise effect in the final fused DEM and reduce the number of PU errors but the improvement is not as large as for the Huber model.  
 
 \subsection{Fusion of TanDEM-X Raw DEMs with Different Baseline Configuration }\label{sec:diffbase}
 
\begin{figure}[htb]
 	\begin{minipage}[b]{0.45\linewidth} 
 		\centering
 		\centerline{\includegraphics[width=1\columnwidth]{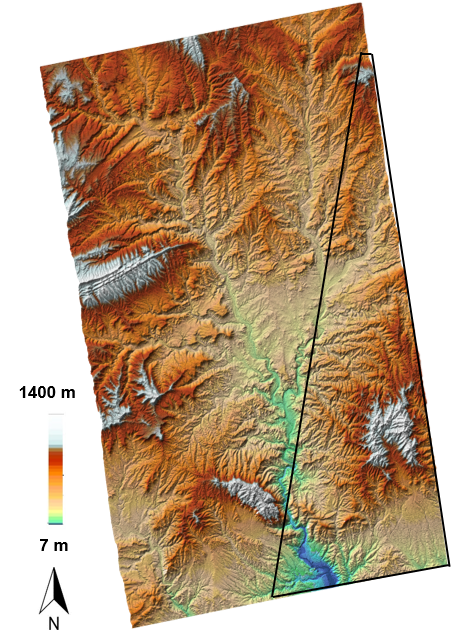}}
 		\vspace{0.05cm}
 		\centerline{(a)}\medskip
 	\end{minipage}
 	\hspace{0.1cm}
 	\begin{minipage}[b]{0.45\linewidth} 
 		\centering
 		\centerline{\includegraphics[width=1\columnwidth]{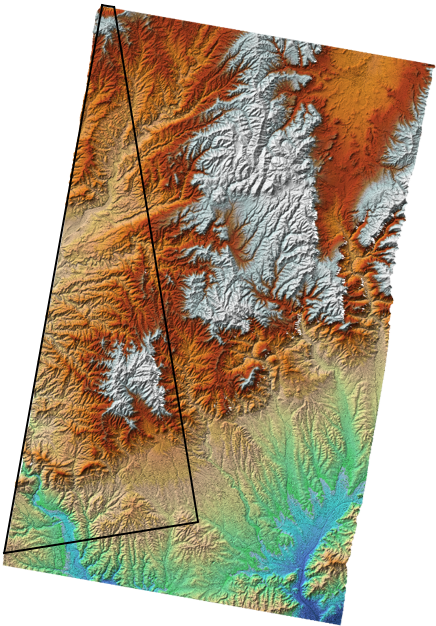}}
 		\vspace{0.05cm}
 		\centerline{(b)}\medskip 
 	\end{minipage}
 	\caption{The ascending (a) and descending (b) tiles  of TanDEM-X raw DEMs produced over  Terrassa and Vacarisses cities}
 	\label{fig:TDXtile-ascdsc}
 \end{figure}

\begin{figure}[htb]
	\begin{minipage}[b]{1\linewidth}
		\centering
		\centerline{\includegraphics[width=1\columnwidth]{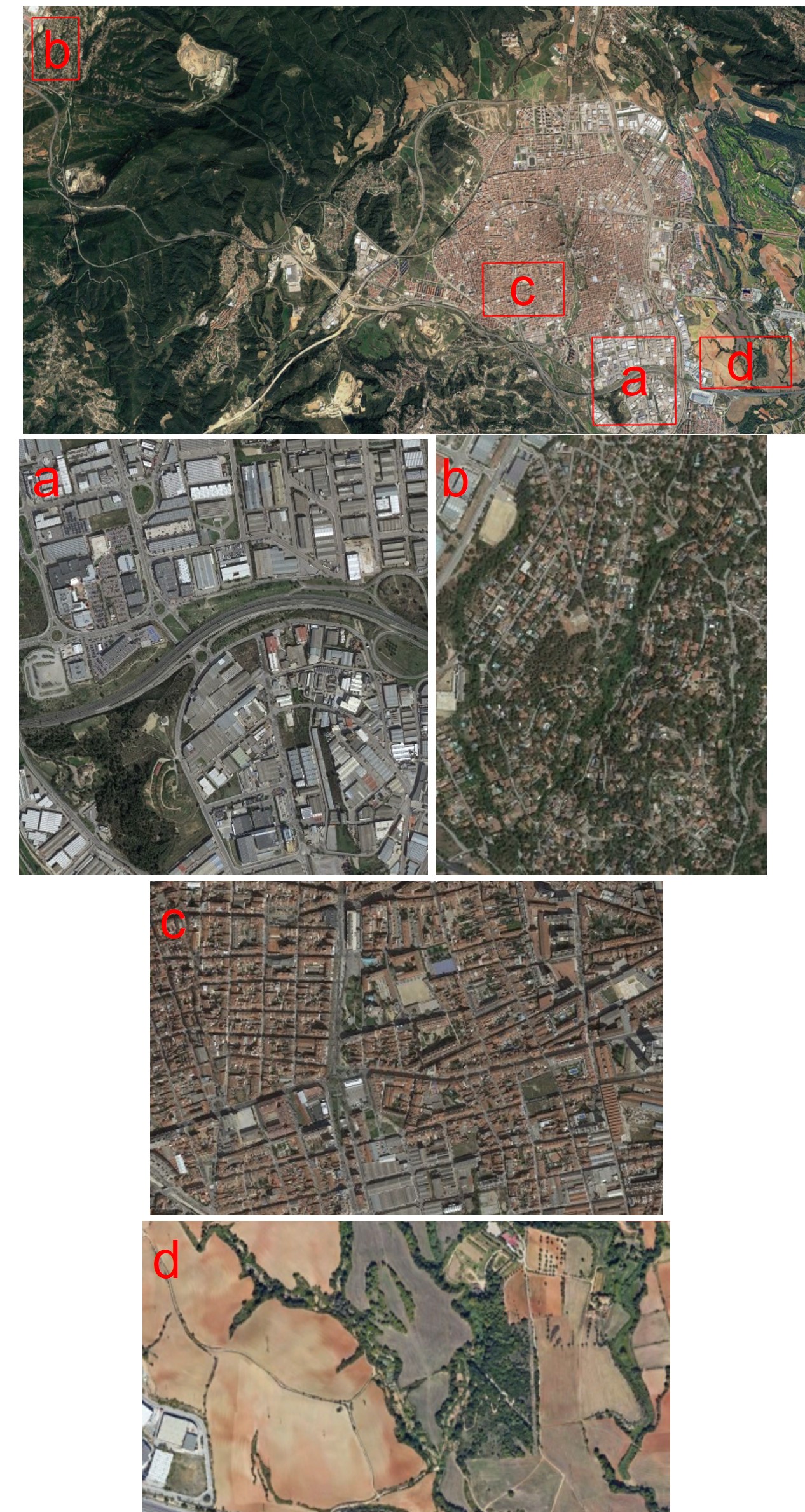}}
		\vspace{0.05cm}
	\end{minipage}    
	\caption{ A display of study subsets selected from different land types for raw TanDEM-X fusion over urban areas; a) industrial area located in Terrassa (1.5km $ \times $ 1.4km), b) residential area located in Vacarisses (1.3km $ \times $0.9km), c) inner city subset located in Terrassa (1km $ \times $ 0.8km), and d) agricultural area (1.5km $ \times $ 0.8km)}
	\label{fig:subset2}
\end{figure}

 In the final experiment, we focus on the fusion of DEMs acquired by different baseline configurations including different orbit directions and HoAs.
 Table \ref{PropertyTDX_ASCDSC} provides the properties of the tiles used for this experiment. The raw DEMs covering Terrassa and Vacarisses cities located in Spain were produced by ascending and descending acquisitions. In addition to orbit directions, the HoAs of tiles are also not similar to each other. Figure \ref{fig:TDXtile-ascdsc} shows the TanDEM-X raw DEMs used in this study which mostly covers  difficult terrain, the common area is specified by black polygons. Due to morphologically difficult type of terrain, the acquisitions from ascending and descending flight paths have been applied for global DEM generation in this area. However, the study cities are located in the relatively flat part of the area common between to tiles. Again, from these tiles study subsets located in different land types were selected. Figure \ref{fig:subset2} display each study subsets from different types extracted from the common area of ascending and descending raw tiles. 
 
 The results of fusing ascending and descending raw DEMs in different land types over urban area are provided in Tab \ref{tab:result4}. The accuracy evaluation is performed by comparing each DEM respective to a LiDAR DSM which was achieved by interpolation of the LiDAR point cloud presented by the ISPRS foundation as a bench mark \cite{ISPRS2018}. On average, the density of the point cloud is about 1 point per square meter. 
 
  \begin{table}[h]
  	\centering
  	\caption{Properties of the nominal ascending and descending TanDEM-X raw DEM tiles over Terrassa and Vacarisses cities.}
  	%\resizebox{\columnwidth}{!}{%
  	\begin{tabular}{l c c }
  		\hline
  		\multicolumn{3}{c}{TanDEM-X raws DEMs}\\
  		Acquisition Id                 & 1058683               & 1171358                     \\\hline
  		Acquisition mode      & Stripmap  & Stripmap \\
  		Center incidence angle      & 33.71$^{\circ}$  & 34.82$^{\circ}$  \\
  		Equator crossing direction  & Ascending        & Descending          \\
  		Look direction              & Right            & Right                  \\
  		Polarization                & HH               & HH                        \\
  		Height of ambiguity         & 60.18 m          & 48.58 m                \\
  		Pixel spacing               & 0.2 arcsec       & 0.2 arcsec        \\
  		HEM mean                    & 1.17 m           & 1.40                   \\
  		\hline		
  	\end{tabular} 
  	%} 
  	
  	\label{PropertyTDX_ASCDSC}
  	
  \end{table}     
 

 	\begin{table*}[h]
 	\centering
 	\caption{Height accuracy (in meter) of the ascending and descending TanDEM-X data before and after DEM fusion in the different study areas over Vacarisses and Terrassa}
 	%\resizebox{\columnwidth}{!}{%
 	\begin{tabular}{lll|ccccc}\hline
 		
 		Study area&\multicolumn{2}{c|}{DEM}& Mean & RMSE & MAE & NMAD & STD \\
 		\hline
 		\multirow{5}{*}{Industrial}  		
		&\multirow{2}{*}{Raw DEM}   &1058683    &-0.19	&3.49	&2.49	&2.60	&3.48   \\
		&                           &1171358    &-0.19	&3.56	&2.44	&2.34	&3.55   \\
		&\multirow{3}{*}{Fused DEM} & WA        &-0.26	&3.06	&2.13	&2.07	&3.05   \\
		&                           & TV-$ L_1 $&-0.34	&2.92	&\textbf{2.09}	&\textbf{2.07}	&2.90   \\
		&                           & Huber	    &\textbf{-0.19}	&\textbf{2.89}	&2.10	&2.14	&\textbf{2.88 }  \\
		\hline
		\hline
 		\multirow{5}{*}{Inner}
 		&\multirow{2}{*}{Raw DEM}  &1058683    &-0.78	&5.05	&3.52	&3.70	&4.99    \\
 		&                          &1171358    &-0.78	&5.11	&3.53	&3.62	&5.05    \\
 		&\multirow{3}{*}{Fused DEM}& WA        &-0.76	&4.66	&3.22	&\textbf{3.36}	&4.59    \\
 		&                          & TV-$ L_1 $&-0.91	&4.35	&\textbf{3.08}	&3.40	&\textbf{4.25}    \\
 		&                          & Huber	    &\textbf{-0.78}	&\textbf{4.34}	&3.13	&3.52	&4.27    \\
 		\hline
 		\hline
 		\multirow{5}{*}{Residential}
 		&\multirow{2}{*}{Raw DEM}   &1058683    &-0.54	&4.24	&3.11	&3.19	&4.20 \\
 		&                           &1171358    &-0.54	&4.42	&3.21	&3.26	&4.38 \\
 		&\multirow{3}{*}{Fused DEM} & WA        &-0.62	&3.94	&2.87	&2.83	&3.90 \\
 		&                           & TV-$ L_1 $&-0.76	&3.96	&2.88	&2.77	&3.88 \\
 		&                           & Huber	    &\textbf{-0.54}	&\textbf{3.86}	&\textbf{2.86}	&\textbf{2.74}	&\textbf{3.82} \\
 		\hline
 		\hline
        \multirow{5}{*}{Agricultural}
        &\multirow{2}{*}{Raw DEM}  &1058683    &0.44	&2.38	&1.68	&1.71	&2.34   \\
        &                          &1171358    &0.44	&1.93	&1.23	&0.98	&1.88   \\
        &\multirow{3}{*}{Fused DEM}& WA        &0.35	&\textbf{1.60}	&\textbf{1.04}	&0.83	&1.57   \\
        &                          & TV-$ L_1 $&\textbf{0.27}	&\textbf{1.60}	&\textbf{1.04}	&\textbf{0.78}	&1.59   \\
        &                          & Huber	    &0.44	&1.62	&1.12	&0.91	&\textbf{1.56}  	\\
        \hline
        
%        \multirow{5}{*}{Mix of land types}
%        &\multirow{2}{*}{Raw DEM}  &1058683    &0.57	&1.77	&1.16	&0.99	&1.68  \\
%        &                          &1171358    &0.57	&1.77	&1.21	&1.15	&1.67  \\
%       &\multirow{3}{*}{Fused DEM}& WA        &0.49	&\textbf{1.27}	&\textbf{0.89}	&\textbf{0.76}	&\textbf{1.17 } \\
%        &                          & TV-$ L_1 $&\textbf{0.47}	&1.56	&1.04	&0.86	&1.49  \\
%        &                          & Huber	   &0.57	&1.39	&0.99	&0.80	&1.27  \\
%         \hline

 	\end{tabular}
 	%	}
 	
 	\label{tab:result4}
 \end{table*}  

The results of DEM fusion again illustrate using variational models can increase the accuracy of the initial input raw DEMs. In urban study subsets, the performance of the Huber model is slightly better than TV-$ L_1 $ according to the statistical metrics but their differences are not really significant. It can be concluded both models produces similar results in terms of statistical measurements. In comparison to WA, variational models also give a more accurate DEM in urban areas and the agricultural subsets.

\section{Discussion}\label{sec:dis}
In this study, the TV-$ L_1 $ and Huber variational models were implemented to fuse TanDEM-X raw DEMs over urban areas as well as surroundings. In particular, we investigated these models with respect to the fusion of raw DEMs produced from data takes with different baseline configurations and HoAs. In conclusion, the results demonstrated the efficiency of variational models in comparison to simple WA for the TanDEM-X raw DEM fusion. To clarify the role of smoothness constraint and data term, we carried out an experiment regarding the DEM quality improvement to be achieved by just carrying out TV-$ L1 $ denoising of a single input DEM. Comparing these results with those achieved by TV-$ L1 $ DEM fusion (which employs elevation data of at least two DEM tiles) revealed that fusion is always favorable (cf. Tab \ref{tab:denoising}). Furthermore, it can be seen that in the industrial subset, the main improvement arises from the smoothness term, which is caused by the regular scene structure. However, adding another tile in a fusion manner can still improve the quality of the final DEM. In contrast, for the agricultural subset, the TV-$ L1 $ denoising could not change the DEM quality, while DEM fusion could finally produce a DEM with higher accuracy. Using more DEM tiles is furthermore vital for areas suffering from layover and shadowing effects or containing phase unwrapping errors. More examples of these areas and requirement for employing several tiles can be found in \cite{7109106}. In addition, regarding the strict quality control policy of the TanDEM-X mission –- obtaining lower than 2m relative height accuracy for slopes lower than 20\% and better than 4m for steeper slopes in each pixel -- means that the pixel-wise TanDEM-X target accuracy can only be realized by DEM fusion. Tab. \ref{tab:denoising} provides some statistics relevant to TanDEM-X quality control indicating percentages of pixels with an accuracy better than 2m and 4m as well as the percentage of pixels with accuracy worse than 4m. The results confirm that DEM fusion can lead to obtaining more reliable pixels in comparison to just the denoising of single DEMs. Another important problem with using a single tile is selection of the accurate subsets in different land types. As an example, in experiment 1, the accuracy of DEM with id 1145180 is higher than the accuracy of DEM 1023491 for the forested area while for other study subsets the quality of DEM 1023491 is better than for the other DEMs. As a result, in practice, it is beneficial to carry out DEM fusion in general, as this always improves the quality of the final DEM.  

	\begin{table*}[h]
	\centering
		\caption{Comparison of TV-$ L_1 $ denoising and TV-$ L_1 $ DEM fusion in industrial and agricultural areas used in the first experiment  }
	%\resizebox{\columnwidth}{!}{%
	\begin{tabular}{ll|cccc}\hline
		
		Study area&Strategy & RMSE & MAE & NMAD  \\
		
		\multirow{2}{*}{Residential}		
		 &TV-$ L_1 $ DEM denoising    & {3.88}	&{2.88}	&{2.20} \\	
		 &TV-$ L_1 $ DEM fusion & \textbf{3.67}	&\textbf{2.69}	&\textbf{2.03}	  \\
		 \hline
		 && Error $ < $ 2m  & Error $ < $ 4m  & Error $ >= $ 4m  \\
		\multirow{2}{*}{Residential}
		 &TV-$ L_1 $ DEM denoising &	49 \%	&78 \%	& 22 \%   \\
		 &TV-$ L_1 $ DEM fusion &	\textbf{54 \%}	&\textbf{81 \%}	&\textbf{19 \%}   \\
		 \hline
		 \hline
		 Study area&Strategy & RMSE & MAE & NMAD  \\
		 \multirow{2}{*}{Agricultural}		
		 &TV-$ L_1 $ DEM denoising    & {0.86}	&{0.57}	&{0.59} \\	
		 &TV-$ L_1 $ DEM fusion & \textbf{0.55}	&\textbf{0.29}	&\textbf{0.20}	  \\
		 \hline
		 && Error $ < $ 2m  & Error $ < $ 4m  & Error $ >= $ 4m  \\
		 \multirow{2}{*}{Agricultural}
		 &TV-$ L_1 $ DEM denoising &	95 \%	&100 \%	& 0 \%   \\
		 &TV-$ L_1 $ DEM fusion &	\textbf{100 \%}	&\textbf{100 \%}	&\textbf{0\%}   \\
		 
	\end{tabular}
	%	}

	\label{tab:denoising}
\end{table*}

\subsection{Use of TV-Based Variational Models}\label{sec:dis-model}

\begin{figure}[htb]	
	\begin{minipage}[b]{0.49\linewidth}
		\centering
		\centerline{\includegraphics[width=1\columnwidth, height=0.8\columnwidth]{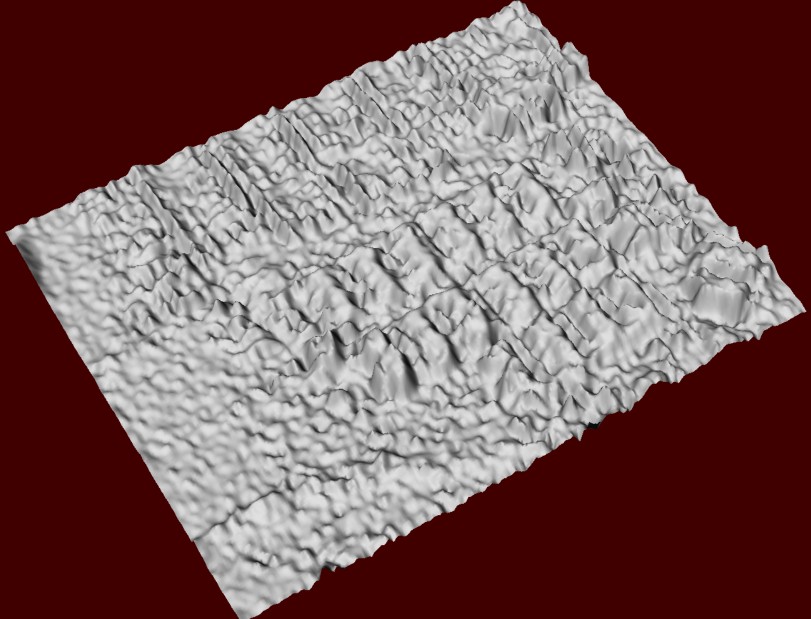}}
		\vspace{0.05cm}
		\centerline{(a) TanDEM-X (tile a) }\medskip
	\end{minipage}
	\begin{minipage}[b]{0.49\linewidth}
		\centering
		\centerline{\includegraphics[width=1\columnwidth, height=0.8\columnwidth]{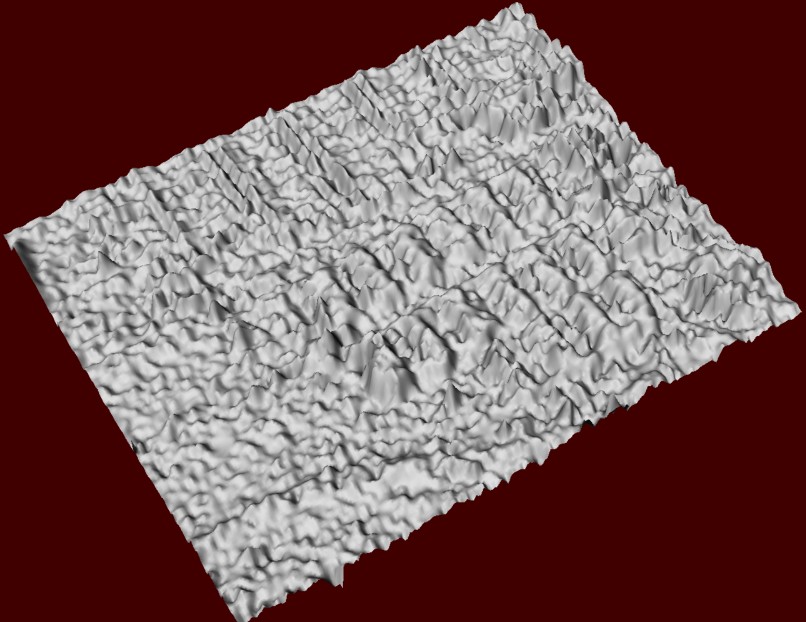}}
		\vspace{0.05cm}
		\centerline{(b) TanDEM-X (tile b)}\medskip
	\end{minipage}
	\begin{minipage}[b]{0.49\linewidth}
		\centering
		\centerline{\includegraphics[width=1\columnwidth, height=0.8\columnwidth]{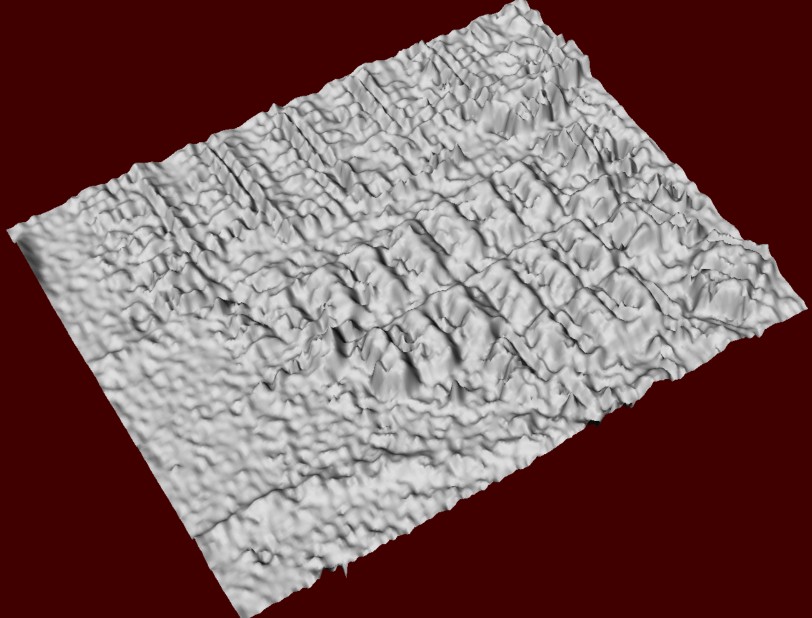}}
		\vspace{0.05cm}
		\centerline{(c) Weighted averaging}\medskip
	\end{minipage}
	\begin{minipage}[b]{0.49\linewidth}
		\centering
		\centerline{\includegraphics[width=1\columnwidth, height=0.8\columnwidth]{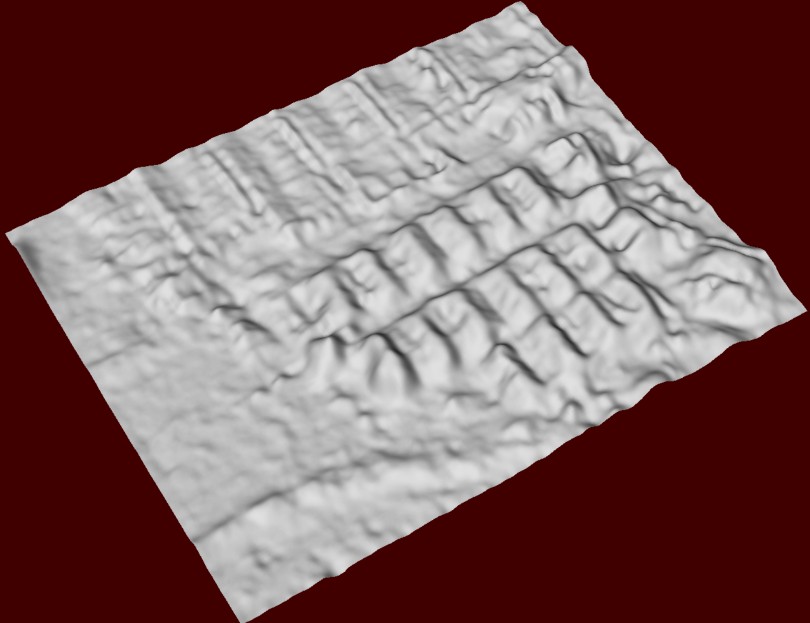}}
		\vspace{0.05cm}
		\centerline{(d) Huber model}\medskip
	\end{minipage}
	\centering
	\begin{minipage}[b]{0.49\linewidth}
		\centering
		\centerline{\includegraphics[width=1\columnwidth, height=0.8\columnwidth]{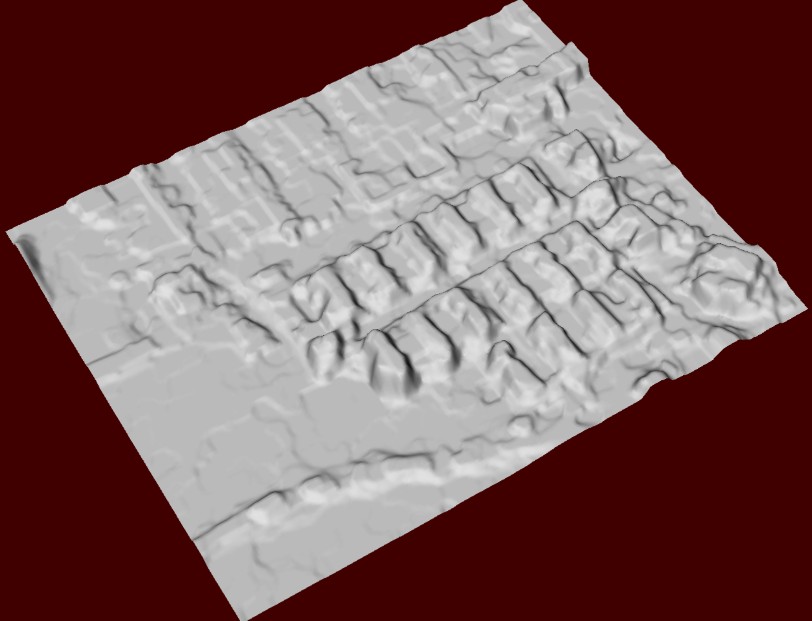}}
		\vspace{0.05cm}
		\centerline{(e) TV-L1 model}\medskip
	\end{minipage}
		\begin{minipage}[b]{0.49\linewidth}
		\centering
		\centerline{\includegraphics[width=1\columnwidth, height=0.8\columnwidth]{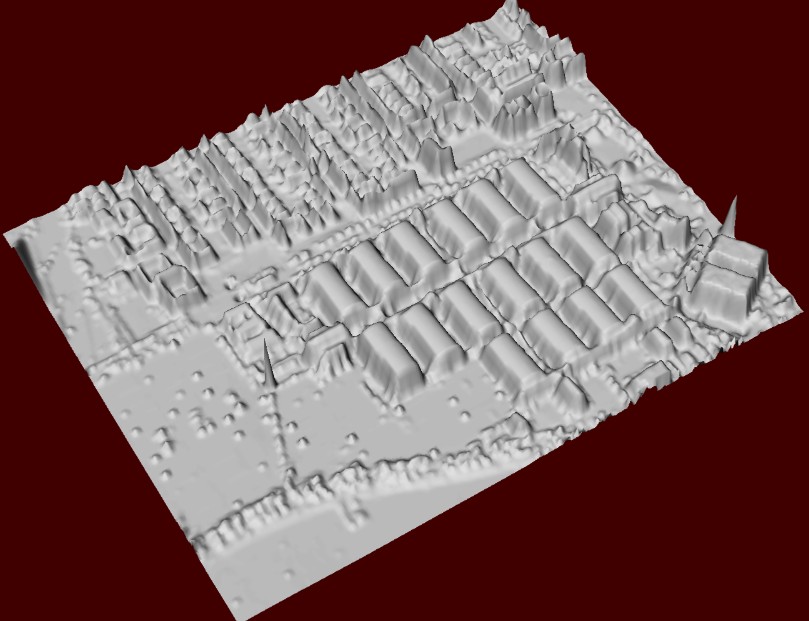}}
		\vspace{0.05cm}
		\centerline{(f) LiDAR }\medskip
	\end{minipage}

	\caption{3D display of initial TanDEM-X raw data and the results of DEM fusions using different methods in the industrial area used in the first experiment.}
	\label{fig:3dfusion}
\end{figure}

The main property of TV-based models is to reduce the effect of noise by minimizing the TV term. It should be noted both data and regularization terms in the energy functional defined for TV-$ L_1 $ and Huber models are positive terms. Choosing TV as a regularization term leads to preserving the beneficial high frequency image contents such as footprints of buildings while minimizing its value through the fusion causes to reduce the effects of undesirable noise. Figure \ref{fig:3dfusion} shows the performance of TV-based variational models in comparison to WA in a 3D view. The displayed patch was selected from an industrial area located in Munich  which was used in the first experiment.  
           
\begin{figure}[htb]
	\begin{minipage}[b]{0.49\linewidth}
		\centering
		\centerline{\includegraphics[width=1\columnwidth, height=0.8\columnwidth]{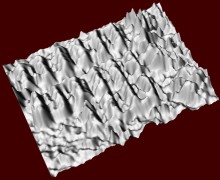}}
		\vspace{0.05cm}
		\centerline{(a) $ \gamma=1 $, $ \alpha=0.5 $, $ \beta=1 $}\medskip
	\end{minipage}
	\begin{minipage}[b]{0.49\linewidth}
		\centering
		\centerline{\includegraphics[width=1\columnwidth, height=0.8\columnwidth]{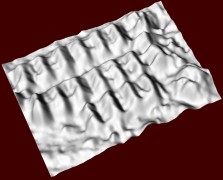}}
		\vspace{0.05cm}
		\centerline{(b) $ \gamma=1 $, $ \alpha=10 $, $ \beta=1 $}\medskip
	\end{minipage}
	\begin{minipage}[b]{0.49\linewidth}
		\centering
		\centerline{\includegraphics[width=1\columnwidth, height=0.8\columnwidth]{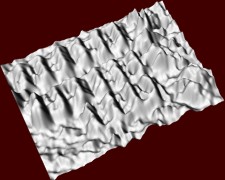}}
		\vspace{0.05cm}
		\centerline{(c) $ \gamma=1 $, $ \alpha=1 $, $ \beta=0.5 $}\medskip
	\end{minipage}
	\begin{minipage}[b]{0.49\linewidth}
		\centering
		\centerline{\includegraphics[width=1\columnwidth, height=0.8\columnwidth]{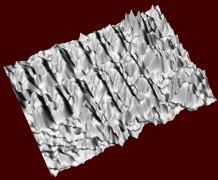}}
		\vspace{0.05cm}
		\centerline{(d) $ \gamma=1 $, $ \alpha=1 $, $ \beta=10 $}\medskip
	\end{minipage}
	\caption{The effect of varying Huber models' parameters on the final DEM.}
	\label{huberparam}
\end{figure}

\begin{figure}[ht!]
	\begin{center}
		\centerline{\includegraphics[width=0.5\textwidth]{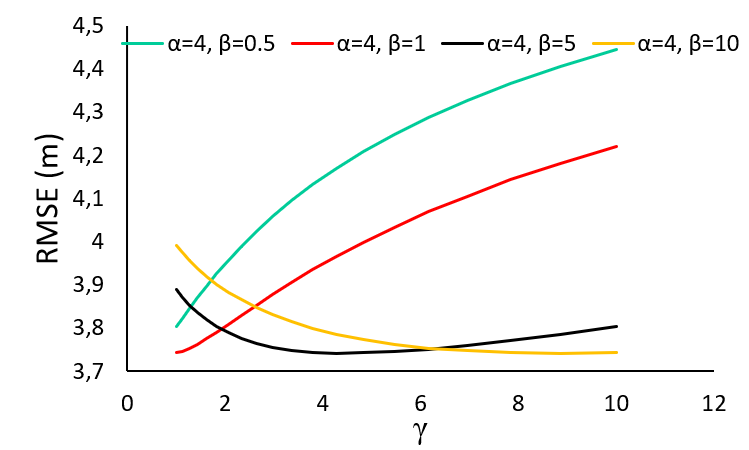}}
		\caption{Influence of different Huber norm parameters on the RMSE of fused DEM}
		\label{fig.hub-tv-gamma}
	\end{center}
\end{figure}

The 3D display of the fused DEMs clearly shows the TV-based model can reduce the noise effect and excellently reveal the edges while the WA-based fused DEM still suffers from noise effects. As displayed, the Huber model produces a smoother output in comparison to TV-$ L_{1} $ because of mixing the quadratic norm and the $ L_{1} $ norm to form data and regularization terms. Since the quadratic norm tends to penalize the high frequency contents more severe than $ L_{1} $, it leads to DEMs with more smoother edges.
Apart from the type of norm used to form an energy functional, the amount of smoothing induced by TV-based variational models depends on the regularization parameter which trades off between the TV term as a regularization term and the data fidelity term. While only one regularization parameter is required to be tuned for DEM fusion by TV-$ L_{1} $, using Huber model for fusion demands to tune 3 parameters.
Selecting different thresholds to form the norms used in the Huber model changes the amount of smoothness that emerges in the final output of DEM fusion. Figure \ref{huberparam} displays the effect of changing one of the parameters while the others are constant on the final output. 
Selecting small $ \alpha $ which is used for data term does not severely penalize discrepancies between the initial DEMs and the desired output strongly i.e. giving an output fused DEM with more similarity to input data. In contrast, increasing $ \alpha $ penalizes the discrepancies intensively and the optimization process tries to lower the total energy that provides a smoother DEM at the end. An identical interpretation can be derived for $ \beta $ while this parameter performs in reverse manner because it is used to form the regularization term. It should be noted the regularization parameter $ \gamma $ trades off between two terms in functional energy that means by increasing $ \gamma $, the effect of TV will become lower such that ultimately smoother DEM is produced.             
Appropriately tuning the regularization parameter and the Huber model thresholds also influences the accuracy of the final fused DEM. The effect of Huber model parameter values on the final accuracy of DEM fusion is depicted in Fig. \ref{fig.hub-tv-gamma}. Different methods can be used for tuning the regularization parameter. One option is to learn it from data if some training data are available. Another option is to use the L-curve approach \cite{doi:10.1137/0914086}. Finally, the parameter can be manually selected based on a visual analysis of different output DEMs. 

%	\begin{figure}[htb]
%		\centering
%%	\begin{minipage}[b]{0.65\linewidth}
%%		\centerline{\epsfig{figure=gamma_1,width=1\columnwidth, height=0.8\columnwidth}}
%%		\vspace{0.05cm}
%%		\centerline{a) $ \gamma $= 0.1}\medskip
%%	\end{minipage}
%%
%%	\begin{minipage}[b]{0.65\linewidth}
%%	\centerline{\epsfig{figure=gamma1,width=1\columnwidth, height=0.8\columnwidth}}
%%	\vspace{0.05cm}
%%	\centerline{b) $\gamma $= 1}\medskip
%%    \end{minipage}
%
%%	\begin{minipage}[b]{0.65\linewidth}
%%	\centerline{\epsfig{figure=gamma10,width=1\columnwidth, height=0.8\columnwidth}}
%%	\vspace{0.05cm}
%%	\centerline{c) $\gamma $= 10}\medskip
%%    \end{minipage}
%
%   \begin{minipage}[b]{1\linewidth}
%	\centerline{\epsfig{figure=huber-gamma,width=0.8\columnwidth, height=0.5\columnwidth}}
%	\vspace{0.05cm}
%	\centerline{  }\medskip
%    \end{minipage}
%	\caption{a,b,c) Influence of different Huber norm parameters $ \alpha $ and $ \beta $ for specific values of the regularization parameter $ \gamma $ on the quality of the fused DEM. d) The relationship between the regularization parameter $ \gamma $ and the RMSE of the fused DEM produced by the Huber model.}
%	\label{fig.hub-tv-gamma}
%\end{figure}          

\subsection{Fusion over Different Land Types}\label{sec:dis-landtype}
\begin{figure}[ht!]
	\begin{center}
		\centerline{\includegraphics[width=1\columnwidth]{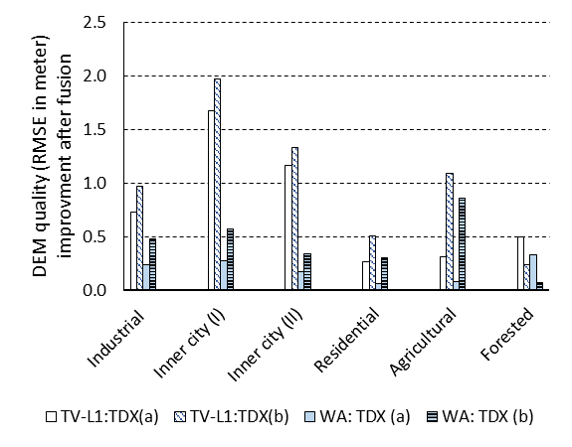}}
		\caption{The improvement of the TanDEM-X DEM tiles (a,b) using variational models (here, TV-L1) in comparison to WA in different study areas located in Munich. The bars indicate the difference between the RMSE of input TanDEM-X DEM and final fused DEM. (a) refers to tile 1023491 and b referes to tile 1145180.}
		\label{improv1}
	\end{center}
	\end{figure}

\begin{figure}[ht!]
	\begin{center}
		\centerline{\includegraphics[width=0.9\columnwidth]{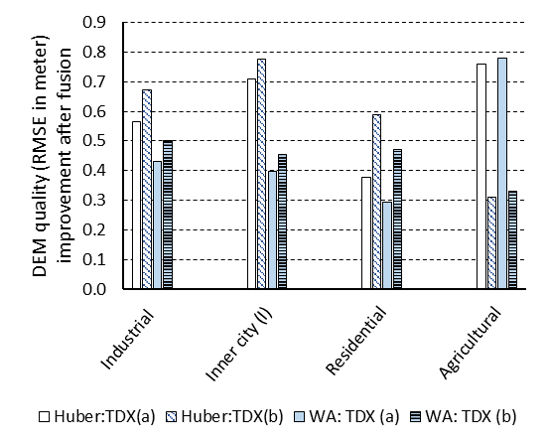}}
		\caption{The improvement of the TanDEM-X DEM tiles (a,b) using variational models (here, Huber) in comparison to WA in different study areas located in Vacarisses and Terrassa. The bars indicate the difference between the RMSE of input TanDEM-X DEM and final fused DEM. (a) refers to tile 1058683 and b referes to tile 1171358.}
		\label{improv2}
	\end{center}
	\end{figure}

In this study, the TanDEM-X DEM fusion by variational models was implemented over different land types that are typically found in urban areas and in their surroundings. Figure \ref{improv1} depicts the accuracy improvement (in meters) by means of different fusion algorithm respective to the quality of the initial DEMs for each study land type used in the first experiment (Section \ref{sec:similarbase}). Similarly, Fig. \ref{improv2} compares the performance of fusion methods for different land types that were used in the third experiment (Section \ref{sec:diffbase}). It should be noted since both variational model have similar performance in terms of RMSE, for each plot in Fig. \ref{improv1} and \ref{improv2}, just performance of the best variational model is compared to WA.

The plots demonstrate that variational models exhibit maximum efficiency in inner city land types in both experiments while WA has a nearly similar performance in different land types. The lowest accuracy improvement by variational models is for residential subset and non-urban study areas. The inner city land type includes a lot of building footprints that mostly appear as noisy edges because of inherent properties of SAR sensor imaging. Consequently, using the TV-based variational model can significantly improve the DEM quality in these areas. On the other hand, residential subsets areas include single family, small homes usually located in a sparse pattern and the footprint of buildings can not appear as strong edge in TanDEM-X raw DEM due to resolution restriction of data takes acquired in stripmap mode. In non-urban areas, the edginess is usually lower than in urban subsets. Thus, the smoothness term of the variational models has lower performance in those kinds of land types. However, the quality of the final DEM still increases due to the DEM fusion encoded in the data term. 

\subsection{The Effect of Geometry}\label{sec:dis-geometry}
While most urban areas covered by global TanDEM-X dataset is generated by two nominal acquisitions which mostly have similar HoAs and geometries, we also investigated the fusion of several TanDEM-X DEMs with different properties to investigate the performance of variational models for these data.
In the first experiment, the results identified that the variational models can perfectly fuse the raw DEMs with nearly similar baseline configuration and HoAs acquired over urban areas. The output is a DEM with higher accuracy and more enhanced building footprints. However, the Huber model generates a smoother DEM at the end. 

A significant result was yielded for problematic areas where the effects of PU errors are dominant. The selected study subset (Fig \ref{fig:studysub2}) is mostly affected by noise because of the volume decorrelation due to trees and the low coherence due to river. For these problematic areas fusing one DEM with nominal HoA to another DEM with larger HoA is more useful to reduce the effect of PU errors. In this experiment, fusing two DEMs with different HoAs by using the Huber model could substitute inconsistent heights with logical values and also resulted in a more accurate DEM. This proves, in addition to DEM fusion methodology, selecting appropriate raw DEM tiles dependent to problem is significant for a successful fusion. Among variational models, TV-$ L_{1}  $ can decrease the number of PU errors and improve the accuracy but more quality enhancement is achieved by the Huber model. Since, the Huber model also uses the quadratic norm, it produces a smoother fused DEM while TV-$ L_{1} $ tends to save more high-frequency contents that can also be caused by noise.  

\begin{figure*}[htb]
		\centering
		
		\begin{minipage}[b]{0.4\linewidth}
			\centerline{\includegraphics[width=1\columnwidth,height=0.8\columnwidth]{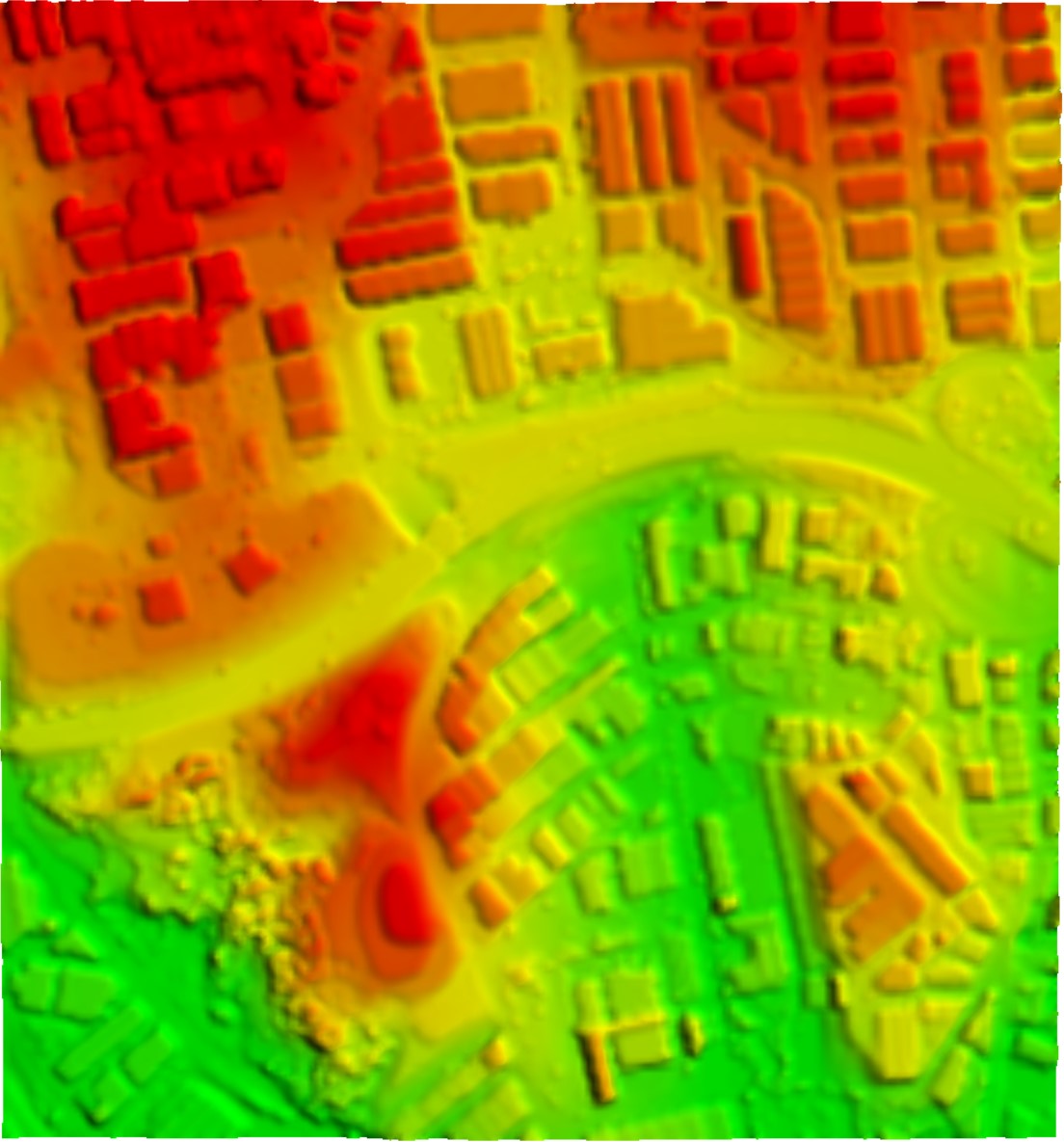}}
			\vspace{0.05cm}
			\centerline{a) LiDAR }\medskip
		\end{minipage}
	\begin{minipage}[b]{0.4\linewidth}
		\centerline{\includegraphics[width=1\columnwidth,height=0.8\columnwidth]{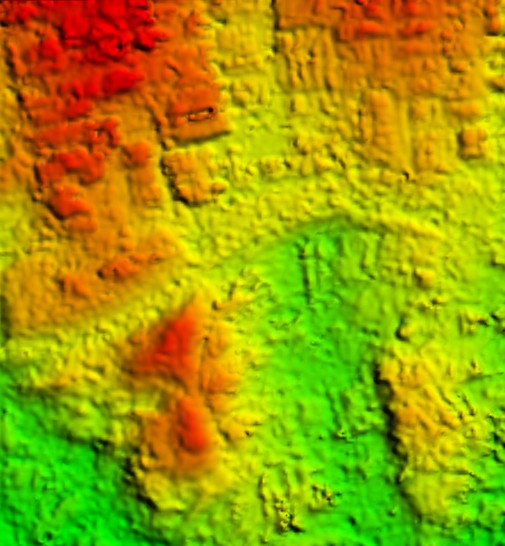}}
		\vspace{0.05cm}
		\centerline{b) WA }\medskip
	\end{minipage}

	\begin{minipage}[b]{0.4\linewidth}
		\centerline{\includegraphics[width=1\columnwidth,height=0.8\columnwidth]{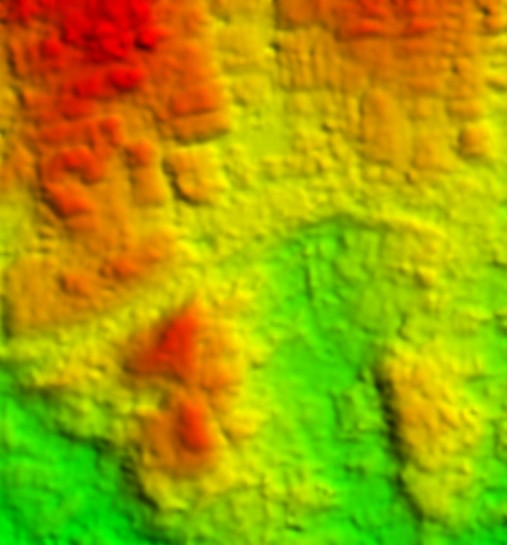}}
		\vspace{0.05cm}
		\centerline{c) Huber}\medskip
	\end{minipage}
	\begin{minipage}[b]{0.4\linewidth}
		\centerline{\includegraphics[width=1\columnwidth,height=0.8\columnwidth]{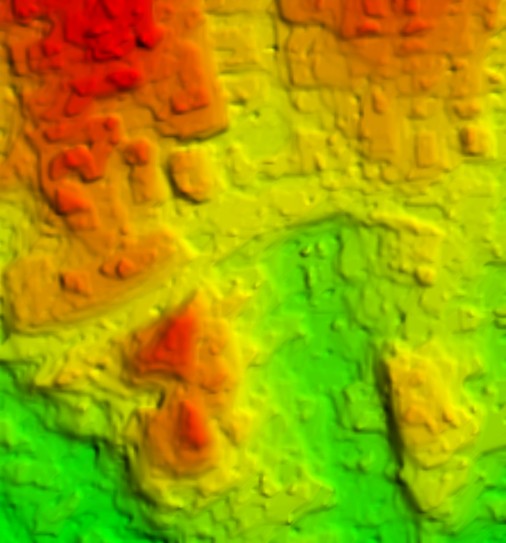}}
		\vspace{0.05cm}
		\centerline{d) TV-$ L_{1} $}\medskip
	\end{minipage}
	\caption{DEMs produced by fusing ascending and descending DEMs over industrial study area located in Terrassa.}
	\label{fig.industS}
\end{figure*}

Fusing ascending and descending DEMs in problematic areas reduces the layover and shadow effects in the final fused DEM. Consequently, in the final experiment, two ascending and descending DEMs with different HoAs were fused.  As shown in plots \ref{improv1} and \ref{improv2}, in comparison to results of fusing DEMs with similar baseline configuration and HoAs, the variational models lead to least significant quality improvement in the final fused DEM in terms of RMSE. However, a display of an exemplary study subset (industrial area) in Fig. \ref{fig.industS} demonstrates the efficiency of variational models in comparison to WA for fusing these types of DEMs. For making a correct judgment about the performance of variational models on fusing ascending and descending DEMs versus DEMs with similar flight paths, two DEMs with similar baseline configuration and HoA from the study areas are required. Theoretically, apart from the DEM fusion method, using ascending and descending DEMs instead of using DEMs with similar orbit directions improves the final DEM quality in the difficult terrains and problematic areas such as urban areas that are under the shadow and layover effects. In practice, it is confirmed in \cite{7109106} that using ascending and descending raw TanDEM-X DEMs can produce highly accurate fused DEM at the end in the shadow- and layover-affected areas.                                     
         
\section{Conclusion}
In this paper, we proposed to apply TV-based variational models (TV-$ L_1 $ and Huber models) for TanDEM-X raw DEM fusion at the phase of DEM  mosaicking instead of weighted averaging. The main focus of this study was to enhance final DEMs in urban areas where the footprints of buildings are influenced by noise effects due to SAR imaging properties. For this purpose, different study subsets were selected from different land types which mostly are explored over urban areas and surroundings. Apart from this, DEM fusion was investigated for raw DEMs with different geometries. At first, two nominal acquisitions with similar baseline configurations and HoAs were fused over different land types. In the next experiment, two raw DEMs with different HoAs were fused over a problematic terrain that suffers from PU errors. At the end, two DEMs with ascending and descending orbit directions as well as with different HoAs were used. In all experiments it was demonstrated that using variational models leads to DEMs with higher quality. A great performance of the Huber model was recorded for fusing two raw DEMs with different HoAs over the selected problematic area. Also, in urban areas, variational models  with reducing the noise effect and enhancing the outlines of buildings, absolutely performs better than WA. However, the Huber model tends to provide a smoother fused DEM than TV-$ L_1 $. The results also demonstrated, the variational models, particularly TV-$ L_1 $, could improve the quality of DEMs significantly in comparison to WA. Using variational models could improve the DEM quality by up to 2m particularly in inner city subsets.    
In conclusion, carrying out TanDEM-X raw DEM fusion using variational models with an ability to enhance the building footprints and other useful high-frequency contents along with smoothing the noise, finally produced a DEM with higher quality.

\section{ACKNOWLEDGMENT}
\label{sec:Acknowlegedment}

The authors would like to thank Dr. Fritz of DLR  and Dr. Baier for providing the TanDEM-X raw DEM for Munich area; and the Bavarian Surveying Administration for providing the LiDAR data of Munich.

\bibliographystyle{ieeetr}
\bibliography{TanDEM-Xfusion}

\begin{IEEEbiography}[{\includegraphics[width=1in,height=1.25in,clip,keepaspectratio]{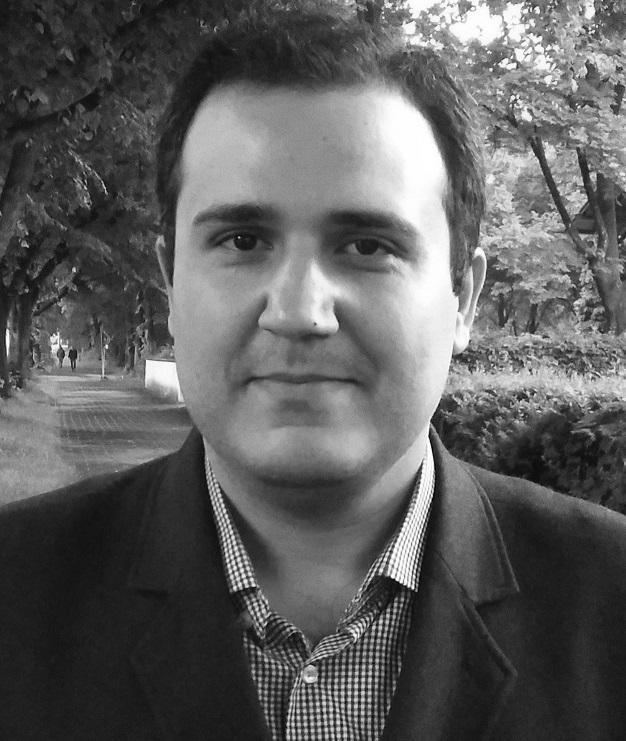}}]{Hossein Bagheri}
Hossein Bagheri received the master's degree in Photogrammetry and Remote Sensing engineering from Tafresh University, Tafresh, Iran, in 2013. He is
currently pursuing the Ph.D. degree with the Technical University of Munich, Munich,
Germany.
In 2015, he spent six months with the Photogrammetry Group, Department of Surveying and Geomatics Engineering, University of Tehran, Iran. His
research interests include multi-sensor data fusion applied to SAR and optical data, 3D reconstruction, and digital elevation models.
\end{IEEEbiography}

\begin{IEEEbiography}[{\includegraphics[width=1in,height=1.25in,clip,keepaspectratio]{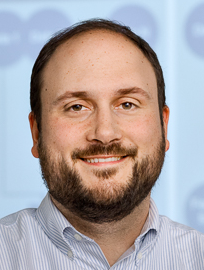}}]{Michael Schmitt}(S'08--M'14--SM'16) received the
Dipl.-Ing. degree in geodesy and geoinformation
and the Dr.-Ing. degree in remote sensing from the
Technical University of Munich (TUM), Munich,
Germany, in 2009 and 2014, respectively.
Since 2015, he has been a Senior Researcher and
the Deputy Head at the Professorship for Signal
Processing in Earth Observation, TUM. In 2016,
he was a Guest Scientist with the University of
Massachusetts Amherst, Amherst, MA, USA. His
research interests include signal and image processing as well as machine learning for the extraction of information from remote sensing data, data fusion with emphasis on the joint exploitation of optical and radar
data, 3-D reconstruction by techniques, such as SAR interferometry, SAR
tomography, radargrammetry, or photogrammetry, and millimeter-wave SAR
remote sensing.
Dr. Schmitt is a Co-Chair of the International Society for Photogrammetry
and Remote Sensing Working Group I/3 on SAR and Microwave Sensing.
He frequently serves as a reviewer for a number of renowned international
journals. In 2013 and 2015, he was elected as the IEEE GEOSCIENCE AND
REMOTE SENSING LETTERS Best Reviewer, leading to his appointment as
an associate editor of the journal in 2016.
\end{IEEEbiography}

\begin{IEEEbiography}[{\includegraphics[width=1in,height=1.25in,clip,keepaspectratio]{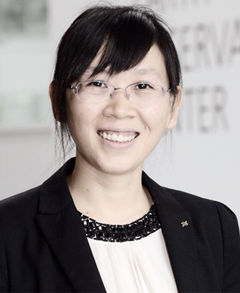}}]{Xiao Xiang Zhu}(S'10--M'12--SM'14) received the Master (M.Sc.) degree, her doctor of engineering (Dr.-Ing.) degree and her “Habilitation” in the field of signal processing from Technical University of Munich (TUM), Munich, Germany, in 2008, 2011 and 2013, respectively.
\par
  She is currently the Professor for Signal Processing in Earth Observation (www.sipeo.bgu.tum.de) at Technical University of Munich (TUM) and German Aerospace Center (DLR); the head of the department ``EO Data Science'' at DLR's Earth Observation Center; and the head of the Helmholtz Young Investigator Group ``SiPEO'' at DLR and TUM. Prof. Zhu was a guest scientist or visiting professor at the Italian National Research Council (CNR-IREA), Naples, Italy, Fudan University, Shanghai, China, the University  of Tokyo, Tokyo, Japan and University of California, Los Angeles, United States in 2009, 2014, 2015 and 2016, respectively. Her main research interests are
  remote sensing and Earth observation, signal processing, machine learning and data science, with a special application focus on global urban mapping.

  Dr. Zhu is a member of young academy (Junge Akademie/Junges Kolleg) at the Berlin-Brandenburg Academy of Sciences and Humanities and the German National  Academy of Sciences Leopoldina and the Bavarian Academy of Sciences and Humanities. She is an associate Editor of IEEE Transactions on Geoscience and Remote Sensing.
\end{IEEEbiography}

\end{document}